\documentclass[a4paper,11pt]{article}
\pdfoutput=1
\usepackage{jheppub}
\usepackage{amsmath}
\usepackage{slashed}
\usepackage{amssymb}
\usepackage{color}
\usepackage{graphicx}
\usepackage{hyperref}
\usepackage{xspace,slashed}
\usepackage[utf8]{inputenc}
\usepackage{subcaption}
\usepackage{multirow}
\usepackage{algorithm}
\usepackage{algorithmic}
\usepackage{stackengine}
\usepackage{enumitem}
\usepackage{etoolbox}

\usepackage{tikz}
\usepackage{tikz-feynman}

\newtoggle{draft}

\toggletrue{draft}

\iftoggle{draft} {
  \usepackage{changes}
}{%
\usepackage[final]{changes}
}

\usepackage{booktabs}

\makeatletter
\g@addto@macro\bfseries{\boldmath}
\makeatother

\newcommand{\defargs}{} 

\newcommand{\barhatN}{{\bar{\hatN}}}
\newcommand{\tildep}{\tilde{p}}

\newcommand{\hatN}{{\bf N}}
\newcommand{\hatS}{{\bf S}}

\newcommand{\ptop}{p_t}

\newcommand{\hatptop}{{\slashed p}_t}
\newcommand{\hatpb}{{\slashed p}_i}

\newcommand{\dtop}{d_t}
\newcommand{\mt}{m_t}
\newcommand{\barmt}{\tilde m_t}

\newcommand{\Gammat}{\Gamma_t}
\newcommand{\qtop}{q_t}

\newcommand{\hatstop}{{\slashed s}_D}
\newcommand{\hatrhotop}{{\slashed \rho}_{t,D}}
\newcommand{\hatetatop}{{\slashed s}_P}
\newcommand{\hatrhotopeta}{{\slashed \rho}_{t,P}}

\newcommand{\pu}{p_u}
\newcommand{\pd}{p_d}
\newcommand{\pb}{p_i}
\newcommand{\pbi}{p_i}

\newcommand{\pfb}{p_f}
\newcommand{\pe}{p_2}

\newcommand{\qe}{q_2}

\newcommand{\tpfb}{{\tilde p}_{f}}
\newcommand{\tpe}{{\tilde p}_{2}}

\newcommand{\qd}{q_d}

\newcommand{\as}{\alpha_s}
\newcommand{\gs}{g_s}

\newcommand\Cf{C_F}

\newcommand{\mathd}{\mathrm{d}}
\newcommand{\tmop}[1]{\ensuremath{\operatorname{#1}}}

\newcommand{\be}{\begin{equation}}
\newcommand{\ee}{\end{equation}}

\usepackage{soul}
\allowdisplaybreaks

\definecolor{azure}{rgb}{0.0, 0.9, 1.0}

\newcommand{\TTPaff}{Institute for Theoretical Particle Physics,
  KIT, 76128 Karlsruhe, Germany}
\newcommand{\MILaff}{INFN, Sezione di Milano-Bicocca, and Universit\`a di Milano-Bicocca, Piazza della Scienza 3, 20126 Milano, Italy}
\newcommand{\Saclay}{Universit\'e Paris-Saclay, CNRS, IJCLab, 91405 Orsay, France}

\preprint{
  \begin {flushright}
    TTP24-028, P3H-24-052
  \end{flushright}
}

\title{
Linear power corrections to single top production and decay at the LHC in the narrow width approximation
}

\author[a]{Sergei Makarov,}
\author[a]{Kirill Melnikov,}
\author[b]{Paolo Nason,}
\author[c]{Melih A. Ozcelik}
\affiliation[a]{\TTPaff}
\affiliation[b]{\MILaff}
\affiliation[c]{\Saclay}

\emailAdd{sergei.makarov@kit.edu}
\emailAdd{kirill.melnikov@kit.edu}
\emailAdd{paolo.nason@mib.infn.it}
\emailAdd{melih.ozcelik@ijclab.in2p3.fr}

\abstract{ We consider top quark production and decay 
in the narrow width approximation  and study
if  the  polarisation effects,
  that manifest themselves in correlations  of angular distributions of  particles  from top quark decays
  and final state  jets  in the production sub-process,  are affected by linear power corrections. We find that, in general, the answer to this question is affirmative.  We also discuss how these non-perturbative corrections 
  affect polarisation observables used to study  
  single top production at the LHC. Finally, we point out that 
  generic   kinematic distributions of leptons 
  from top quark decays are affected by linear power corrections, which may have implications for proposals to extract the top quark mass from such leptonic observables.  On the other hand, we  demonstrate  that the distribution of  the ``out-of-collision-plane'' component of the positron momentum is free from linear power corrections,  making it 
  an interesting candidate for the top quark mass measurement. 
   }

\begin{document}

\maketitle 
\section{Introduction}

In a recent series of papers~\cite{Makarov:2023ttq,Makarov:2023uet} we have  studied 
${\cal O}(\Lambda_{\rm QCD})$ power 
corrections to top quark 
production in hadron collisions using the approach 
based on infra-red renormalons.\footnote{
An in-depth discussion of infra-red renormalons in QCD 
can be found in  review~\cite{Beneke:1998ui}.
For a   detailed description of how these methods can be 
used in the context 
of hadron collider applications, see ref.~\cite{FerrarioRavasio:2018ubr}.}   
In this paper  we  extend these analyses by accounting
for top quark decays.  Such 
an extension is non-trivial. 
Indeed,   since  the top quark width $\Gammat$ serves as an 
infra-red regulator,  its  interplay with the
non-perturbative QCD parameter $\Lambda_{\rm QCD}$ 
and its proxy in the context of renormalon calculus -- the gluon mass $\lambda$ -- is important.  Since  $\Gammat  \gg \Lambda_{\rm QCD} \sim \lambda $,  top quarks decay before hadronisation.  A renormalon-based analysis of power corrections in such a  case  is 
 technically very challenging, 
because  the produced 
top quarks  are off-shell, and diagrams where real or virtual 
gluons  connect production 
and decay stages of   off-shell top quarks have to be considered. Although this problem represents  an 
interesting challenge for future work, we believe that it makes 
sense to start by considering 
the \emph{opposite} case   $\Gammat \ll \Lambda_{\rm QCD} \sim \lambda$, which 
can be studied in the   narrow width approximation. Even  if the  phenomenological relevance
of such an analysis is limited, it provides, for the very 
first time, an estimate of the 
non-perturbative corrections 
to a full  physical process with unstable particles at  a hadron collider and, as such, might 
be quite valuable for modelling the non-perturbative 
effects. 

The narrow width approximation leads to important technical simplifications since, in this case, QCD radiative corrections cannot connect top quark production and top quark 
decay sub-processes \cite{Fadin:1993kt,Melnikov:1993np}.
In fact, the QCD corrections to top quark production and 
decay process are known through next-to-next-to-leading 
order in this  approximation \cite{Berger:2016oht,Campbell:2020fhf}.
However, even when 
the narrow width approximation is used, 
the  production and decay sub-processes 
are not independent; the  
communication between them  occurs because of momentum conservation and \emph{polarisation effects}.\footnote{
Theoretical studies of spin correlations in top quark pair 
production and polarisation effects in single top 
production have a long history, see e.g. refs~\cite{
Kane:1991bg, Aguilar-Saavedra:2014eqa, Mahlon:1996pn, Mahlon:1999gz,Baumgart:2012ay} for original and more recent work.} For the case of single
top production, that we study in this paper, this implies that e.g. the direction of the outgoing light-quark
jet and the direction of the positron in the top quark decay are not independent in spite of the fact that they originate  from 
widely separated sub-processes.
As we will show below,  linear power 
corrections affect  angular distributions in single top production. 
In fact, we find that  they vanish when 
 positrons  from top decays  and light jets from the production process are collinear to each other, but 
 are non-trivial functions of momenta of final state 
 particles otherwise.    We will also 
show that the ${\cal O}(\Lambda_{\rm QCD})$ effects that we discuss  in this paper impact  the various observables 
designed to study top quark polarisation  
\cite{CMS:2013rfa,Komm:2014fca,ATLAS:2022vym} differently, so that  for each of them a dedicated study is required. 

The paper is organised as follows. In the  next section we describe the narrow 
width approximation and present the result for the differential cross section of single top production followed by top decay in a way that is  
useful for the subsequent analysis. In sections~\ref{sect3} and \ref{sect4} we  discuss 
the calculation of ${\cal O}(\lambda)$ corrections 
to the production and decay sub-processes. In 
section~\ref{sect:mass},  ${\cal O}(\lambda)$ corrections due to mass-parameter redefinition are studied. In section~\ref{sect:kinematics}, corrections to observables are discussed and  the final formulas are derived.  We conclude in 
section~\ref{sect:concl}.  In the appendix we illustrate
an alternative approach to the calculation of power corrections where, at  variance with the method used 
in the previous publications \cite{Makarov:2023ttq,Makarov:2023uet}, we deal directly 
with the amplitude of the process rather than with its 
square.

\section{The narrow  width approximation}
\label{sec:narrowwidth}

\begin{figure}[htb]
\begin{center}
\includegraphics[width=0.4\textwidth]{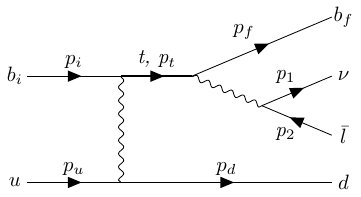}
\caption{The Born amplitude for
single top production and decay.}
\label{fig:Born}
\end{center}
\end{figure}

In this paper we 
consider the following partonic process
\be
\begin{split}
u(p_u) + b(p_i) \to d(p_d) + & t(p_t) \\
&~~~~ {}^{\searrow} \; b(p_f)+  \nu(p_1) + e^+(p_2).
\label{eq2.1}
\end{split} 
\ee
It is shown  in Fig.~\ref{fig:Born}.
The amplitude for this process can be  written as
\be
{\cal M} = A_{\rm dec}^{i} \frac{ i \delta^{ij} (\hatptop + \mt)}{p_t^2 - \mt^2 + i \mt \Gammat}
  A_{\rm prod}^{j},
\ee
where $i,j$ are colour indices.  In the on-shell  $p_t^2 \to \mt^2$ limit,
the quantities $A_{\rm dec}$ and $A_{\rm prod}$ correspond to on-shell  amplitudes for the respective  sub-processes
from which the top quark spinors are  removed.

The expression for the cross section becomes
\be
   {\rm d} \sigma_{PD} = \frac{1}{\mathcal{N}} \frac{ {\rm d}\Phi(p_u,p_{i}; p_d,\{p_{\rm dec} \})  }{(p_t^2 - \mt^2)^2 + \mt^2 \Gammat^2 } \;
    \sum \limits_{\rm spins} A_{\rm dec}^{i}(\hatptop + \mt) A_{\rm prod}^{i} {\bar A}^{j}_{\rm prod} (\hatptop + \mt) {\bar  A}^{j}_{\rm dec},
  \label{eq2.3a}
   \ee
   where the suffix $PD$ indicates production and decay, ${\rm d} \Phi$ is the 
   phase space of the process
   in eq.~(\ref{eq2.1}),
   the sum over spins includes external particles only and $\{p_{\rm dec} \}$ describes the momenta $p_{f, 1, 2}$ that refer to particles 
   originating  from the  ``decay'' of the  virtual top quark.   The normalisation factor $\mathcal{N}$ includes all the averaging factors needed
to compute the cross section of the process in eq.~(\ref{eq2.1}) as well as the relevant flux factor.    

To simplify eq.~(\ref{eq2.3a}), we note that since in the narrow width approximation no colour transfer between production
and decay amplitudes occurs, the following equation holds 
\be
A_{\rm dec}^{i}...{\bar  A}^{j}_{\rm dec} = \frac{\delta^{ij}}{N_c} A_{\rm dec}^{k}...{\bar  A}^{k}_{\rm dec},
\ee
where $N_c=3$ is the number of colours.
Hence,
\be
   {\rm d} \sigma_{PD} = \frac{1}{\mathcal{N} \; N_c} \frac{ {\rm d}\Phi(p_u,p_{i}; p_d,\{p_{\rm dec} \})  }{(p_t^2 - \mt^2)^2 + \mt^2 \Gammat^2 } \;
   \sum \limits_{\rm spins} A_{\rm dec}^{i}(\hatptop + \mt) A_{\rm prod}^{j} {\bar A}^{j}_{\rm prod} (\hatptop + \mt) {\bar  A}^{i}_{\rm dec}.
\label{eq2.3b}
\ee
To proceed further, we factorise the 
phase space
\be
   {\rm d}\Phi(p_u,p_i;p_d,\{p_{\rm dec} \} ) =
   \frac{{\rm d} p_t^2}{2 \pi} \; {\rm d}\Phi(p_u,p_i;p_d,p_t) \; {\rm d}\Phi(p_t;\{ p_{\rm dec} \} ),
   \ee
and make use of the fact that we work in the narrow width approximation which implies that 
the following equation holds
\be
\frac{1}{(p_t^2 - \mt^2)^2 + \mt^2 \Gammat^2 } \Bigg |_{\Gammat/\mt \to 0} = \frac{2 \pi}{2 \mt \Gammat} \delta(p_t^2 - \mt^2).
\label{eq:narrowwidth}
\ee
  We obtain
\be
   {\rm d} \sigma_{PD} = \frac{ {\rm d}\Phi(p_u,p_i;p_d,p_t) \; {\rm d}\Phi(p_t;\{ p_{\rm dec} \} ) }{\mathcal{N} \; N_c \ 2 \mt  \Gamma_t }  \;
  \sum \limits_{\rm spins} A_{\rm dec}^{i}(\hatptop + \mt) A_{\rm prod}^{j} {\bar A}^{j}_{\rm prod} (\hatptop + \mt) {\bar  A}^{i}_{\rm dec}.
\label{eq2.3}
\ee

It is easy to see that  the following  (matrix) equation holds\footnote{This is due to the fact that helicities of all other external particles are fixed, so that the top quark  must be in a pure spin state. See appendix for more details.}
\be
\sum \limits_{\rm spins} (\hatptop + \mt)  {\bar  A}^{i}_{\rm dec} A_{\rm dec}^{i}(\hatptop + \mt)
= X \, \hatrhotop,
\label{eq2.9}
\ee
where the sum extends over the polarisations 
of the top quark decay products and where we have defined the top quark spin density matrix
\begin{equation}
    \hatrhotop = ( \hatptop + \mt) \frac{\left ( 1 + \gamma_5 \hatstop \right )}{2}.
    \label{eq2.10a}
\end{equation}
In eq.~(\ref{eq2.9}), $X$  is a function of scalar products constructed out of the top quark momentum $p_t$ and the momenta of its   decay products, 
and 
$s_D^\mu$ is a space-like unit vector 
that also depends  on $p_t$ and the momenta of the particles in top decay.  To find $X$, we compute the trace of both sides of eq.~(\ref{eq2.9})
and obtain
\be
\sum \limits_{\rm spins} 2 \mt  {\rm Tr} \left [ (\hatptop + \mt )  {\bar  A}^{i}_{\rm dec} A_{\rm dec}^{i} \right ] = 2 \mt  \; X.
\ee
We can write this equation in a better way by  using the formula for the differential decay width of the unpolarised top quark
\be
   {\rm d} \Gammat  = \frac{{\rm d}\Phi(p_t;\{ p_{\rm dec} \})  }{4 \mt N_c } \; \sum \limits_{\rm spins}
   {\rm Tr} \left [ (\hatptop + \mt)  {\bar  A}^{i}_{\rm dec} A_{\rm dec}^{i} \right ].
\ee
It follows that
\be
X = \sum \limits_{\rm spins} {\rm Tr} \left [ (\hatptop + \mt)  {\bar  A}^{i}_{\rm dec} A_{\rm dec}^{i} \right ]
 = 4 \mt N_c \; \frac{{\rm d} \Gamma_t}{{\rm d} \Phi(p_t;\{p_{\rm dec} \})}.
\ee
Finally,  using this result as well as eq.~(\ref{eq2.9}), we re-write eq.~(\ref{eq2.3}) as 
\be
   {\rm d} \sigma_{PD}  = 2 \frac{ {\rm d} \Gammat}{\Gammat } \times {\rm d} \sigma_{t}(s_D).
\label{eq2.13}
   \ee
In the above equation,  ${\rm d} \sigma_t(s_D)$ is the cross section for producing a single top quark 
whose spin is aligned with the 
axis $s_D$;
it is defined as
\be
   {\rm d} \sigma_{t}(s_D) = \frac{ {\rm d}\Phi(p_u,p_i;p_d,p_t)}{\mathcal{N}}  \;
  \sum \limits_{\rm spins} {\rm Tr} \left [ \hatrhotop A_{\rm prod}^{j} {\bar A}^{j}_{\rm prod}\right].
\ee
We note that we will refer to  the spin  axis $s_D$ as the top quark spin vector below.  
It is important
to emphasise that  for 
the process in eq.~(\ref{eq2.1})
this vector depends on the momenta of  the top quark decay products.  It is computed in the appendix  and given in eq.~(\ref{eq:sD}).
The calculation of the top quark polarised cross section in eq.~(\ref{eq2.13}) 
proceeds in the standard way. The only difference with the unpolarised case
is that instead of the density matrix $(\hatptop + \mt)$ one has to use
$\hatrhotop$.

The formula for the cross section shown in eq.~(\ref{eq2.13}) is suitable for computing  QCD corrections to the production
process followed by  tree-level decay.
When 
tree-level production is followed by the
QCD-corrected decay, it is  more convenient  to deal with the decay
of the polarised  top quark.  Following the discussion above, 
we find
\be
   {\rm d} \sigma_{PD} = 2 \frac{ {\rm d} \Gammat(s_P)}{\Gammat } \times {\rm d} \sigma_{t},
\label{eq2.16}
   \ee
   where this time ${\rm d} \sigma_t$ is the unpolarised top quark production cross section and ${\rm d} \Gamma_t(s_P)$ is the decay rate of a polarised
   top quark
   \be
   {\rm d} \Gammat {(s_P)}  = \frac{{\rm d}\Phi(p_t;\{ p_{\rm dec} \})  }{4 \mt N_c } \; \sum \limits_{\rm spins}
   {\rm Tr} \left [ \hatrhotopeta  {\bar  A}^{i}_{\rm dec} A_{\rm dec}^{i} \right ].
   \ee
   The corresponding spin density matrix reads
   \begin{equation}
    \hatrhotopeta = ( \hatptop + \mt) \frac{( 1  
    + \gamma_5 \hatetatop  )}{2}.
\end{equation}
   The polarisation vector of the top quark $s_P$ is given in eq.~(\ref{eq:sP}).
 We note that at  leading order 
 the two representations of the full differential cross section, given in eqs~(\ref{eq2.13}) and (\ref{eq2.16}),  are  equivalent, so that the following equation holds
 \be
{\rm d} \sigma_t  \; {\rm d}  \Gammat(s_P) = {\rm d} \sigma_t(s_D) \; {\rm d} \Gammat.
\label{eq2.18}
\ee
This equation allows us to  use one representation 
 to study radiative corrections to the production process and another one to study radiative 
 corrections to  top decay.  Furthermore,  eq.~(\ref{eq2.18})
 will be  useful for studying effects related to the redefinition of the top quark mass. Such a redefinition    is needed 
 to  remove  ${\cal O}(\lambda)$ corrections  caused by the fact that the pole mass of the top quark is used in perturbative
 computations  and that this mass parameter itself 
 receives ${\cal O}(\lambda)$ corrections when expressed through a short-distance mass \cite{Bigi:1994em,Beneke:1994sw}.

Eq.~(\ref{eq2.18}) will also be useful for simplifying the final result. In particular, for single top production, the above 
equation assumes a particularly simple form\footnote{For the derivation of this formula, see eq.~(\ref{eq:bpdsq}) and the discussion before it.}
\be
{\rm d} \sigma_t \; {\rm d}  \Gamma_t(s_P) = {\rm d} \sigma_t(s_D) \; {\rm d} \Gamma_t 
 = \frac{1}{2} {\rm d} \sigma_t {\rm d} \Gamma_t 
 \left ( 1 - s_D \cdot s_P \right ).
 \label{eq2.19}
\ee
 Hence, by choosing the explicit 
 form of one of the two spin vectors and  pretending that the other one is general,  one  can put  the polarisation 
 information either to the  production or to the decay 
 sub-process of the full process 
 in eq.~(\ref{eq2.1}).

\section{QCD corrections to the production sub-process}
\label{sect3}

We consider the  production sub-process 
\begin{equation}
  u(\pu) + b(\pbi) \to d (\pd) + t(\ptop),
  \label{eq1.1a}
\end{equation}
with the assumption that the top quark is polarised.
As explained in 
ref.~\cite{Caola:2021kzt}, corrections to the 
light-quark line do not 
produce  linear ${\cal O}(\lambda)$ contributions.  For this reason, we only need to 
 consider the QCD corrections to the heavy-quark line. 
To describe this
process, we can use eq.~(\ref{eq2.13}) and compute the standard perturbative contributions to the polarised cross section.
In our discussion, we will assume that the  reader is familiar with the computation for the unpolarised case reported in 
ref.~\cite{Makarov:2023ttq}, and 
we will mostly emphasise the 
differences between the two cases in what follows.

\subsection{Real emission contributions}
\label{sect:real}
We begin with the real emission  corrections to the 
process in eq.~(\ref{eq1.1a})
\begin{equation}
 u(\pu) + b(\pbi) \to d (\qd) + t(\qtop) + g(k).
\label{eq8.1}
\end{equation}
The gluon $g(k)$ is massive, $k^2 = \lambda^2$.
We remark that we have used slightly different notations for the four-momenta  of the top quark and the down quark in eqs~(\ref{eq1.1a}) and (\ref{eq8.1}).
This is done for 
future convenience  since, as we will see, 
momenta $q_t$ and $q_d$ will absorb the recoil  due to  the emitted soft gluon. We will also consider the top quark to be polarised, and we will denote its
spin vector with $s_D^\mu$.

The calculation proceeds along the lines described  in  ref.~\cite{Makarov:2023ttq} where the case
of stable top quark was discussed.    The main element 
of  that discussion was the 
Low-Burnett-Kroll theorem 
\cite{Low:1958sn,Burnett:1967km} that can be used 
to describe soft radiation in QCD with 
next-to-leading-power accuracy.\footnote{An extension of this theorem to processes with polarised particles can be found in ref.~\cite{Fearing:1973eh}.} 
This  theorem was derived from 
the transversality 
of the amplitude with respect to the  gluon momentum
which  allowed us to relate  soft gluon emission 
by the  external particles and  the structure-dependent radiation.  
The fact that
the top quark spinor represents  a state with a particular polarisation plays no role in this argument. 
Hence, upon  writing
\begin{equation}
  {\cal A}_{\rm prod} = \gs T^{a}_{ij} \epsilon_\mu {\cal M}^\mu,
\end{equation}
and repeating all the steps
described in ref.~\cite{Makarov:2023ttq}, we arrive at the following result
\begin{equation}
\begin{split} 
   {\cal M}^\mu & =
   J_t^\mu \bar u_t \hatN(\qtop+k,\pb,\qd,..) u_i + J_i^\mu \bar u_t \hatN(\qtop,\pb-k,\qd,..) u_i
   \\
   & + \bar u_t  \left [ \hatS_t^{\mu} \hatN(\qtop,\pb,\qd,..) + \hatN(\qtop,\pb,\qd,..) \hatS_i^\mu \right ] u_i
   \\
&    - \bar u_t \left [ \frac{  \partial \hatN(\qtop,\pb,\qd,...)}{ \partial q_{t, \mu}} +  \frac{\partial
       \hatN(\qtop,\pb,\qd,...)}{\partial p_{i, \mu}}  \right ]  u_i.
   \end{split}
\label{eq3.4}
\end{equation}
In the above expression, $\hatN$ is the tree-level  amplitude for single top production from which top and bottom spinors have been
removed, and  
\begin{equation}
  \begin{split} 
    & J_t^\mu = \frac{2\qtop^\mu + k^\mu}{\dtop},\;\;\;\; \hatS_t^\mu = \frac{\sigma^{\mu \nu} k_\nu}{\dtop},
    \\
& J_i^\mu = \frac{2 \pb^\mu  - k^\mu}{d_i},\;\;\;\; \hatS_i^\mu = \frac{\sigma^{\mu \nu} k_\nu}{d_i},
  \end{split} 
  \label{eq:JtJb}
  \end{equation}
are top (bottom) currents and spin operators, respectively. The two quantities in the above  
equation, 
$d_t = (q_t + k)^2 - \mt^2$ and $d_i = (p_i - k)^2$, are the denominators  of top and bottom  propagators.
Furthermore, $\sigma^{\mu \nu} = [\gamma^\mu,\gamma^\nu]/2$.

We can further simplify the expression in eq.~(\ref{eq3.4}) by combining the first two terms, expanded to first subleading power in $k$, with the last two terms. We obtain
\begin{equation}
\begin{split}
   {\cal M}^\mu & =
   J^\mu \bar u_t \hatN(\qtop,\pb,\qd,..) u_i 
   +   \bar u_t ( L^\mu\hatN(\qtop,\pb,\qd,..) ) u_i
   \\
   & + \bar u_t  \left [ \hatS_t^{\mu} \hatN(\qtop,\pb,\qd,..) + \hatN(\qtop,\pb,\qd,..) \hatS_i^\mu \right ] u_i,
\end{split}
\label{eq8.15}
\end{equation}
where we have introduced the notation
\begin{equation}
J^\mu = J_t^\mu + J_i^\mu,\;\;\;\;L^\mu = L_t^\mu - L_i^\mu,
\label{eq:JLdef}
\end{equation}
with 
\begin{equation}
L_t^\mu = \frac{2}{d_t}\left(p_t^\mu k^\nu \frac{\partial }{\partial \qtop^\nu} - (p_t k)\frac{\partial }{\partial q_{t, \mu}}\right),
\;\;\;\;
L_i^\mu = \frac{2}{d_i}\left(p_i^\mu k^\nu \frac{\partial }{\partial p_i^\nu} -(p_i k)  \frac{\partial }{\partial p_{i,\mu}}\right).
\label{eq:Lbdef}
\end{equation}

Eq.~(\ref{eq8.15}) expresses the amplitude with the emission of a soft gluon through the elastic amplitude and its derivatives. However, we observe further simplifications  if we square the amplitude and sum over the polarisations of the external particles, or consider external states with definite  helicities. We find
\begin{align}
& |{\cal M}|^2 = -g_{\mu \nu} {\cal M}^\mu {\cal M}^{\nu,+}
  = -J^\mu J_\mu  {\rm Tr} \left [ \hatrhotop  \hatN \hatpb {\barhatN}\defargs \right ] \nonumber 
        \\
& - J^\mu {\rm Tr} \left [ \hatrhotop  \hatN \hatpb L_\mu    {\barhatN}\defargs  \right ]
-J_\mu  {\rm Tr} \left [    \hatrhotop ( L^\mu \hatN ) 
  \hatpb  {\barhatN}\defargs  \right ]
\label{eq3.10}
\\
&
+J_\mu {\rm Tr} \left [ [\hatS^\mu_t,\hatrhotop] \hatN \hatpb {\barhatN} \right ] +
J_\mu {\rm Tr} \left [ \hatrhotop  \hatN [\hatpb,\hatS^\mu_i ] {\barhatN} \right ].
\nonumber 
\end{align} 
Note that the only difference between eq.~(\ref{eq3.10}) and a similar expression for $|{\cal M}|^2$  in ref.~\cite{Makarov:2023ttq}
is that the density matrix $\hatrhotop$ defined in 
eq.~(\ref{eq2.10a}) appears in  eq.~(\ref{eq3.10}) instead of $(\hatptop+m_t)$. 
Since,
\begin{equation}
   [\hatS_t^\mu, \hatrhotop ] = \left (  -L_t^\mu - S_t^\mu \right )  \hatrhotop = -(L^\mu + S_t^\mu)  \hatrhotop,
   \;\;\;\;\; [\hatS_i^\mu,\hatpb] = -L_i^\mu \hatpb =  (L^\mu + S_t^\mu) \hatpb,
   \label{eq3.11}
\end{equation}
where
\be
S_t^\mu = \frac{2}{d_t}\left( s_D^\mu k^\nu \frac{\partial }{\partial s_D^\nu} - ( s_D k) \frac{\partial }{\partial s_{D,\mu}}\right),
\label{eq:Stdef}
\ee
we obtain
\begin{equation}
\begin{split}
  &  |{\cal M}|^2   = -J^\mu J_\mu {\rm Tr} \left [ \hatrhotop  \hatN \hatpb {\barhatN}\defargs \right ] 
- J^\mu {\rm Tr} \left [ \hatrhotop \hatN \hatpb L_\mu {\barhatN}\defargs  \right ]
-J^\mu  {\rm Tr} \left [  \hatrhotop  ( L_\mu \hatN ) 
  \hatpb  {\barhatN}\defargs  \right ]
\\
&
-J^\mu {\rm Tr} \left [ (( L_\mu +S_{t,\mu}) \hatrhotop ) \hatN \hatpb {\barhatN} \right ]
-J^\mu {\rm Tr} \left [ \hatrhotop   \hatN ( (L_\mu  +S_{t,\mu}) \hatpb) {\barhatN} \right ].
\end{split}
\end{equation}
We emphasise that in the above formulas, starting from  eq.~(\ref{eq3.11}), derivatives with respect 
to the four-momenta of partons that appear in the 
operators $L_{t/i}$ do not act on the polarisation vector $s_D^\mu$.

Making use of the fact that $L^\mu$ is a linear differential operator, and that the only dependence
on $s_D^\mu$ is in the density matrix $\hatrhotop$, we combine
the last four terms to obtain a derivative 
of the leading order  polarised amplitude. The final result reads 
\begin{align}
& |{\cal M}|^2 
  = - \left ( J^\mu J_\mu  + J_\mu ( L^\mu +S_t^\mu)  \right ) F_p(p_u,p_i,q_t,q_d,s_D),
\label{eq2.22}
\end{align}
where
\be
F_p( p_u,p_i,q_t,q_d,s_D) =  {\rm Tr} \left [ \hatrhotop \hatN \hatpb {\barhatN}\defargs \right ],
\label{eq3.14a}
\ee
is the matrix element squared for the production process with polarised top quark and 
where, by construction, derivatives  with respect to momenta do not act on $ \hatstop$. We also note that one can obtain 
the unpolarised result by simply  
setting $s_D \to 0$ in eq.~(\ref{eq3.14a}) 
and multiplying the result by a factor $2$. By a slight abuse 
of notation, we will write 
\be
2 F_p(p_u,..,q_d,s_D \to 0) \equiv F_p(p_u,..,q_d), 
\ee
in what follows.

Similarly to the stable-top case, the first term on the right hand side
in eq.~(\ref{eq2.22}) contributes to the top quark production cross section starting at order  ${\cal O}(\lambda^0)$, whereas
the second and the third ones contribute at order ${\cal O}(\lambda)$.   To  extract ${\cal O}(\lambda)$ contributions from the first term 
in eq.~(\ref{eq2.22}),  we  redefine the momenta
of various particles to remove the momentum $k$ from the energy-momentum conserving 
delta-function.  However,
  an important difference between stable and unstable top cases  occurs at this point because the top quark momentum appears
  both in the production and in the decay phase spaces. \emph{Hence, redefinition of the top quark momentum in the production process
  leads to the redefinition of the top quark momentum in  the decay and one needs to understand how to deal with it.}
\\

To adhere as much as possible to the discussion of single top production and top decay that  were studied separately in ref.~\cite{Makarov:2023ttq},  
we employ the   momenta redefinitions used 
there.  We  write 
\begin{equation}
\begin{split}
& q_t = p_t - k + \frac{(p_t  k)}{ (p_t   \pd)}  \pd,\;\;\;\; \qd = \pd - \frac{(p_t k)}{ (\ptop \pd)}  \pd.
\end{split} \label{eq:momentamapping}
\end{equation}
Repeating the steps described in ref.~\cite{Makarov:2023ttq}, we find 
\begin{equation}
  {\rm d}\Phi_P(\pu,\pbi;\qd,q_t,k) = {\rm d}\Phi_P(\pu,\pbi; \pd, p_t) \; [{\rm d} k]_\lambda 
 \; \left ( 1 + \frac{(k \pd) }{ (p_t \pd)} - \frac{(k p_t)}{ (p_t \pd)} 
 +{\cal O}(\lambda^2)
 \right ),
 \label{eq3.17}
\end{equation}
where
\be
[{\rm d} k]_\lambda = \frac{{\rm d}^4 k}{(2\pi)^3} \delta_+(k^2 - \lambda^2).
\ee

In the stable-top case, once the above transformation is performed and relevant matrix elements squared are expanded in $k$,
integration over $k$ becomes possible. The same happens when decay is considered except that we need to account
for the change in the differential decay width and the decay phase space introduced by 
momenta redefinitions in eq.~(\ref{eq:momentamapping}).

Momenta redefinitions are only relevant for 
the leading ${\cal O}(\lambda^{-2} ) $ term in  eq.~(\ref{eq2.22}). Its contribution to the cross section is proportional to 
\be
\begin{split} 
   & {\rm d}\Phi_P(\{q_{\rm in}\}; \pd, \ptop) \;  
      \;  {\rm d} \Gammat(q_t,\{q_{\rm dec} \}) \; J^{(0)}_\mu J^{(0),\mu} \;
  F_p(..,q_t,q_d, s_D(q_t,q_2))|_{q_t \to p_t+ .., q_d \to p_d+..},
\label{eq3.23}
\end{split}
\ee
where ${\rm d} \Phi_P$ is the phase space of the 
production subprocess after momenta redefinition 
and expansion in $k$, and $J^{(0)}_\mu$ is the leading 
power contribution to the eikonal current given 
in eq.~(\ref{eq:JLdef}).
The required momenta shifts are shown in eq.~(\ref{eq:momentamapping}). We note that the differential width that appears in the above expression depends on  the \emph{original} top quark momentum $q_t$. Furthermore, momenta redefinitions also affect  the spin vector of the top quark $s_D$ as it depends on the top quark momentum  $q_t$.

To understand how to expand eq.~(\ref{eq3.23}) to 
first sub-leading order in the gluon momentum,
 we note  that the top quark momentum redefinition 
 in eq.~(\ref{eq:momentamapping}) can be interpreted as a Lorentz transformation. 
 Indeed, we can write eq.~(\ref{eq:momentamapping}) as follows 
\be
q_t^\mu = \Lambda^{\mu \nu} p_{t,\nu},
\ee
where
\be
\label{eq:Lambdak}
\Lambda^{\mu \nu} = g^{\mu \nu} - \frac{k^\mu p_d^\nu - p_d^\mu k^\nu}{ (p_t p_d)}
 = g^{\mu \nu} + \delta \Lambda^{\mu \nu}.
\ee
Since $\delta \Lambda^{\mu \nu}$ is an anti-symmetric traceless matrix, it can be interpreted as   an infinitesimal Lorentz transformation.   

This observation has important implications for the calculation of the differential decay width ${\rm d} \Gammat(q_t,\{q_{\rm dec} \})$ that appears in eq.~(\ref{eq3.23}). Indeed, after momentum 
transformation, it becomes 
\be
{\rm d} \Gammat(q_t,\{q_{\rm dec} \} )= {\rm d} \Gammat(\Lambda p_t,\{q_{\rm dec} \} )
= {\rm d} \Gammat(\Lambda p_t,\{ \Lambda p_{\rm dec} \})
= {\rm d} \Gammat(p_t,\{ p_{\rm dec} \}),
\ee
where we also transformed momenta of the final 
state particles in the decay and made use 
of the fact that the differential width 
is invariant under Lorentz transformations. 

Although  the above result implies that momenta 
redefinitions in the production do not generate 
${\cal O}(k)$ corrections 
in the unpolarised decay width that 
appears in eq.~(\ref{eq3.23}),  the need to redefine momenta of final-state particles in the 
decay has implications for the spin vector $s_D$. 
Indeed, $s_D$  
depends on $q_t$ and $q_2$ and, therefore, changes when the above momenta transformations are performed.  It is easy to see that this change is described by a Lorentz boost 
\be
s_D^\mu(q_t,q_2) = \Lambda^{\mu \nu} s_{D,\nu}(p_t,p_2),
\label{eq3.23a}
\ee
where $\Lambda^{\mu \nu}$ is given in 
eq.~(\ref{eq:Lambdak}).

We are now in a position to write the result for the ${\cal O}(\lambda)$ contribution  to the   differential cross section of the process 
in eq.~(\ref{eq2.1}) due to an emission of a soft 
gluon in the production sub-process.  It  arises as a sum of the  correction to the production  
matrix element described in eq.~(\ref{eq2.22}), 
correction to the production phase-space 
shown in eq.~(\ref{eq3.17}) and corrections 
to the leading order term shown in eq.~(\ref{eq3.23}) and discussed afterwards. 
Many of these contributions do not involve modifications of the spin vector $s_D$ and are 
\emph{identical} to  contributions 
studied in 
ref.~\cite{Makarov:2023ttq}.
New spin-dependent 
contributions arise because of the spin-operator 
$S_t$ in eq.~(\ref{eq2.22}) and because of the Lorentz boost of the spin-vector in eq.~(\ref{eq3.23a})
that has to be inserted into the function 
$F_p$ in  eq.~(\ref{eq3.23}) and expanded in $k$. 

Hence, we write 
\be
   {\cal T}_\lambda \left [
    {\rm d} \sigma_{PD}^{R,\rm prod}  \right ]= 2 \frac{{\rm d} \Gammat}{\Gammat} \;  {\cal T}_\lambda\left [ {\rm d} \sigma^R_{t}(s_D) \right ]_k
  + {\cal T}_\lambda \left [    {\rm d} \sigma^{R,\rm prod}_{PD,\rm~ new}  \right ]_k,
  \label{eq3.24}
\ee
where  subscript $k$ indicates that the integration  over gluon momentum is still to be performed. Furthermore, the   first term on the right hand side 
in eq.~(\ref{eq3.24}) 
is \emph{the same as in the no-decay case} \cite{Makarov:2023ttq} except that one employs 
the polarised cross section for single top production, and the second term is new. It reads 
\be
 {\cal T}_\lambda \left [    {\rm d} \sigma^{R,\rm prod}_{PD,\rm new}  \right ]_k
= - \frac{ {\rm d}\Phi_P(... ;p_t,p_d) }{\mathcal{N}}
[{\rm d} k ]_\lambda \;2 \; \frac{{\rm d} \Gammat}{\Gammat} \; J^{(0)}_\alpha \; {\cal O}^{(s),\alpha} \; 
    F_p(...,p_t,p_d,s_D(p_t,\pe)),
\ee
   where 
   \be
 {\cal O}^{(s),\alpha} = 
 S_t^\alpha - J^{(0),\alpha}  s_{D,\mu} \; \delta \Lambda^{\mu \nu} \frac{\partial}{\partial s_D^\nu}.
   \ee

To complete the computation of the real-emission 
contributions we need to integrate over the gluon 
momentum.  This is straightforward since   in both old and new contributions  in eq.~(\ref{eq3.24})
the dependence on the gluon momentum $k$ is exposed
and one can use the formulas 
in appendix A of ref.~\cite{Makarov:2023ttq} 
to calculate the relevant integrals over $k$. 
Hence, we find 
\be
   {\cal T}_\lambda \left [
    {\rm d} \sigma_{PD}^{R,\rm prod}  \right ]= 2 \frac{{\rm d} \Gammat}{\Gammat} \;  {\cal T}_\lambda\left [ {\rm d} \sigma^R_{t}(s_D) \right ]
  + {\cal T}_\lambda \left [    {\rm d} \sigma^{R,\rm prod}_{PD,\rm  new}  \right ].
\label{eq3.27}
\ee
The first term on the right hand side 
of eq.~(\ref{eq3.27}) can be found 
in eq.~(2.31) of ref.~\cite{Makarov:2023ttq}, 
where $F_{\rm LO}$ should be replaced with 
$F_p(...,s_D)$,  and the second (new) term reads 
   \be
  {\cal T}_\lambda \left [    {\rm d} \sigma^{R,\rm prod}_{PD,\rm~ new}  \right ] 
         = \frac{\alpha_s C_F}{2 \pi } \frac{\pi \lambda}{\mt} \; 2 \; \frac{ {\rm d} \Gammat}{ \Gammat}\;
         s_D^\mu  \; W_{\mu \nu}  \;
         \frac{\partial}{\partial s_{D,\nu}} \; {\rm d} \sigma_t(s_D).
         \; 
                         \label{eq3.32}
   \ee
   The rank-two  tensor in eq.~(\ref{eq3.32}) 
   can be written as 
   \be
   W^{\mu \nu} =  \omega^{\mu \nu}_{dt} - \omega^{\mu \nu}_{i t} + \frac{2 m_t^2 (\pbi p_d)}{(p_t \pbi)(p_t p_d) }
   \omega^{\mu \nu}_{id},
   \ee
where the quantity 
   \be
\omega_{xy}^{\mu \nu} = \frac{ p_x^\mu p_y^\nu - p_y^\mu p_x^\nu}{(p_x p_y)},
\label{eq:omegaxy}
   \ee
describes an infinitesimal boost in the $(p_{x}-p_{y})$ plane.    Eq.~(\ref{eq3.32}) provides  an additional real emission 
contribution  to the ${\cal O}(\lambda)$ terms  
when single top  production process 
   is combined with top quark decay  in the narrow width approximation.

\subsection{Virtual corrections and renormalisation  in the production sub-process}
\label{sect:virt}

As explained in ref.~\cite{Makarov:2023ttq}, the
${\cal O}(\lambda)$ contributions  from the 
virtual corrections can be treated in the same 
way as the real emission ones.  In this section 
we study the virtual corrections to the production 
sub-process of the process in eq.~(\ref{eq2.1}) and 
use the polarised top-quark spinor to describe the  influence of the top quark decay. 

Similarly to the real emission case, the ${\cal O}(\lambda)$ contributions to the virtual corrections can only arise from the region of soft $k \sim \lambda$ loop momenta. Our goal, therefore, is to establish a similar soft expansion of one-loop virtual corrections to the single top production process $u(\pu) + b(\pbi) \to d(\pd) + t(\ptop)$.
We again consider corrections to the heavy-quark
line since corrections to the light-quark line 
do not produce ${\cal O}(\lambda)$ contributions \cite{Caola:2021kzt}. 
We write
\begin{equation}
{\cal A}_{\rm virt} = \gs^2 \Cf \delta_{ij} {\cal M}_{\rm virt},
\end{equation}
where $i,j$ are the colour indices of the top quark and the bottom quark. Proceeding as in ref.~\cite{Makarov:2023ttq},
we find 
\begin{equation}
\begin{split}
  & {\cal M}_{\rm virt} = \int \frac{{\rm d}^4 k}{(2 \pi)^4} \frac{-i}{k^2 - \lambda^2}
  \Bigg [ J_t^\alpha J_{i,\alpha} \; \bar u_t \left (  \hatN(\ptop,\pb,..) + k^\mu D_{p,\mu} \hatN(\ptop,\pb,..)\right ) u_i
     \\
&    - J_t^\alpha \bar u_t  \hatN(\ptop,\pb,..) \hatS_{i,\alpha} u_i
    + J_i^\alpha \bar u_t \hatS_{t,\alpha}  \hatN(\ptop,\pb,..) u_i
   -(J_t^\alpha +J_i^\alpha) \bar u_t D_{p,\alpha} \hatN u_i 
    \Bigg ],
  \end{split}
\end{equation}
where $d_t = (p_t + k)^2 - m_t^2$ and $d_i = (p_i+k)^2$,
\be
D_p^\mu = \frac{\partial }{\partial p_{t,\mu}} + \frac{\partial }{\partial p_{i,\mu}},
\ee
and 
\begin{equation}
  \begin{split} 
& J_t^\alpha = \frac{2 \ptop^\alpha + k^\alpha}{\dtop},\;\;\;\hatS_t^\alpha = \frac{\sigma^{\alpha \beta} k_\beta}{\dtop}, \\
& J_i^\alpha = \frac{2 \pb^\alpha + k^\alpha}{d_i},\;\;\;\hatS_i^\alpha = \frac{\sigma^{\alpha \beta} k_\beta}{d_i}.
\end{split}
\end{equation}

Similar to the real emission case, the dependence on the loop momentum has been made explicit so that the
integration over $k$ becomes straightforward.  However, it is better  to square the matrix element
 before integrating over the loop momentum $k$.  We do this following ref.~\cite{Makarov:2023ttq} and accounting  for the fact that
the top quark is polarised.   We find
\begin{equation}
\begin{split}
& \delta_{\rm virt}[ {\cal M} {\cal M}^+ ]
= \int \frac{{\rm d}^4 k}{(2 \pi)^4} \frac{-i}{k^2 - \lambda^2}
\Bigg [
  2 J_t^\alpha J_{i,\alpha} {\rm Tr} \left [ \hatrhotop \hatN  \hatpb \barhatN \right ]
  \\
&   + J_t^\alpha J_{i,\alpha} k^\mu \; {\rm Tr} \left [ \hatrhotop (D_{p,\mu} \hatN) \hatpb \bar {\hatN} +
    \hatrhotop \hatN \hatpb (D_{p,\mu}   {\bar {\hatN}} )\right ]
  \\
  & -(J_t^\alpha + J_i^\alpha) {\rm Tr} \left [ \hatrhotop ( D_{p,\alpha}\hatN ) \hatpb \barhatN
        + \hatrhotop  \hatN \hatpb ( D_{p,\alpha} \barhatN )
    \right ]
\\
& +  J_i^\alpha {\rm Tr} \left[ [\hatrhotop ,\hatS_{t,\alpha} ]\hatN \hatpb \barhatN \right ]
- J_t^\alpha {\rm Tr} \left[ \hatrhotop \hatN [\hatS_{i,\alpha},\hatpb] \barhatN \right ]
\Bigg  ],
\end{split}
\label{eq3.35a}
\end{equation}
where ${\cal M}={\cal M}_0+{\cal M}_{\rm virt}$. 
The above  equation contains all  ${\cal O}(\lambda)$ 
corrections to $ {\cal M} {\cal M}^+ $.

We can further simplify eq.~(\ref{eq3.35a})  following steps already discussed in the previous section 
where the  real emission contribution was  considered.  Indeed, using 
\begin{equation}
\begin{split} 
  &  [\hatrhotop,
       \hatS_t^\alpha] = ( L_t^\alpha+S_t^\alpha) \hatrhotop,
  \\
    &  [\hatpb, \hatS_i^\alpha] =   L_i^\alpha \hatpb,
\end{split}
    \end{equation}
we arrive at 
    \begin{equation}
\begin{split}
& \delta [ {\cal M}{\cal M}^+ ]_{\rm virt}
= \int \frac{{\rm d}^4 k}{(2 \pi)^4} \frac{-i}{k^2 - \lambda^2}
\Bigg [
  2 J_t^\alpha J_{i,\alpha} {\rm Tr} \left [ \hatrhotop \hatN  \hatpb \barhatN \right ]
  \\
&   + J_t^\alpha J_{i,\alpha} k^\mu \; {\rm Tr} \left [ \hatrhotop (D_{p,\mu}\hatN) \hatpb \barhatN +
    \hatrhotop \hatN \hatpb (D_{p,\mu}   {\barhatN}) \right ]
  \\
  & -(J_t^\alpha + J_i^\alpha) {\rm Tr} \left [ \hatrhotop  ( D_{p,\alpha}\hatN ) \hatpb \barhatN
        + \hatrhotop \hatN \hatpb ( D_{p,\alpha} \barhatN )
    \right ]
\\
& +  J_i^\alpha {\rm Tr} \left[ (  ( L_{t,\alpha} +S_{t,\alpha} ) \hatrhotop )\hatN  \hatpb \barhatN \right ]
 + J_t^\alpha {\rm Tr} \left[ \hatrhotop \hatN ( L_{i,\alpha} \hatpb)  \barhatN \right ]
 \Bigg  ].
\label{eq10.14}
\end{split}
    \end{equation}
    We stress that,  similar to the real emission 
    case, derivatives with respect to momenta 
    do  not act on the 
    spin vector $s_D$ that appears in the density 
    matrix $\hatrhotop$.

    We note that the difference between the result 
    in eq.~(\ref{eq10.14}) 
    and a similar result  computed for the  unpolarised
    case in ref.~\cite{Makarov:2023ttq}, 
is the appearance of the spin-dependent density 
matrix $\hatrhotop$  everywhere in eq.~(\ref{eq10.14})
and  the presence of an additional  term  
that contains the operator $S_{t,\alpha}$. This term evaluates to 
\begin{equation}
\int \frac{{\rm d}^4 k}{(2 \pi)^4} \frac{-i}{k^2 - \lambda^2}\;
J_i^\alpha  \; S_{t,\alpha}  {\rm Tr} \left[ \hatrhotop \hatN  \hatpb \barhatN \right ] = - \frac{1}{8 \pi^2 } \frac{\pi \lambda}{m_t} \; \frac{(p_i s_D)}{(p_i p_t)}p_t^\nu \frac{\partial}{\partial s_D^\nu} \; {\rm Tr} \left[ \hatrhotop \hatN  \hatpb \barhatN \right ],
\end{equation}
as follows from the integrals collected in appendix A 
of ref.~\cite{Makarov:2023ttq}. Hence, we can write the virtual contribution as
\begin{equation}
   {\cal T}_\lambda \left [
    {\rm d} \sigma_{PD}^{V,\rm prod}  \right]= 2 \frac{{\rm d} \Gammat}{\Gammat} \;  {\cal T}_\lambda\left [{\rm d} \sigma^V_{t}(s_D) \right ]
  + {\cal T}_\lambda \left [    {\rm d} \sigma^{V,\rm prod}_{PD, \rm new}  \right ],
  \label{eq3.39}
\end{equation}
where the first term on  the right hand side can be found 
in eq.~(3.16) of ref.~\cite{Makarov:2023ttq} and 
the second term is new.  We note that in the first term  we again need to replace 
$F_{\rm LO}$ with $ F_p(...,s_D)$ 
and $(\hatptop + m_t)$ with $\hatrhotop$ to account for the polarisation effects. 
The second term in eq.~(\ref{eq3.39}) is a new 
contribution. It reads 
\begin{equation}
   {\cal T}_\lambda \left [    {\rm d} \sigma_{PD,\rm new}^{V,\rm prod}  \right ]
         = -\frac{\alpha_s C_F}{2 \pi } \frac{\pi \lambda}{\mt} \; 2 \; \frac{ {\rm d} \Gammat}{ \Gammat} \; 
         s_{D, \mu}  \; \omega^{\mu \nu}_{i t}  \;
         \frac{\partial}{\partial s_{D}^{\nu}} \; {\rm d} \sigma_t(s_D).
         \label{eq3.41}
\end{equation}
Finally,  contributions due to  wave function 
and  (explicit) mass renormalisation are not 
affected by the fact that the top quark 
is polarised. Hence, we conclude that these 
contributions can be borrowed from ref.~\cite{Makarov:2023ttq} without any modification
\begin{equation}
   {\cal T}_\lambda \left [
    {\rm d} \sigma_{PD}^{\rm Ren,\rm prod}  \right]= 2 \frac{{\rm d} \Gamma_t}{\Gamma_t} \;  {\cal T}_\lambda\left [{\rm d} \sigma^{\rm Ren}_{t}(s_D) \right ].
\label{eq3.42}
\end{equation}

To summarise, ${\cal O}(\lambda)$ contributions 
to the cross section of the process 
in eq.~(\ref{eq2.1}) caused by the radiation of real and virtual gluons and the renormalisation  in the 
\emph{production} sub-process are obtained as 
the sum of the  contributions 
given 
in eqs~(\ref{eq3.27}), (\ref{eq3.39}) and  (\ref{eq3.42}).
Each of  these contributions is  written as the sum of two terms:  the ``old 
one'' that are identical to the stable-top
production case discussed in 
ref.~\cite{Makarov:2023ttq}, except for the 
fact that 
one has to employ there the polarised leading 
order cross section, and the ``new one'' which 
is  entirely due to the fact that there are spin correlations between production and decay processes.   When single top production was considered in isolation, ``old corrections'' were 
cancelling against the  redefinition of the top quark 
mass parameter;  a similar cancellation also 
exists in the current case.  However, before discussing 
this point, we need to compute the ${\cal O}(\lambda)$ power correction to the decay sub-process. We do this in the 
next section.

\section{Corrections to the decay sub-process}
\label{sect4}

In this section we explain how the  ${\cal O}(\lambda)$
power correction to the  top quark decay sub-process is  computed. 
Following the discussion in section~\ref{sec:narrowwidth}, the top 
quark is polarised and its polarisation vector 
$s_P$ is determined by the kinematics of the 
production sub-process.   We note that corrections 
to the decay of 
an unpolarised top quark 
considered in isolation can be 
found in appendix B in ref.~\cite{Makarov:2023ttq}.

Our starting point is eq.~(\ref{eq2.16}).  The momenta assignments differ from the ones in appendix B of ref.~\cite{Makarov:2023ttq}; for this reason we emphasise that we consider the decay
process
\be
t(p_t) \to \nu(p_1) + e^+(p_2) + b(p_f). 
\ee

The calculation of the real-emission contributions proceeds similarly to the case of the single top production and along the lines of 
appendix B of ref.~\cite{Makarov:2023ttq}.
We use the following momenta assignment 
to describe the real-emission process 
\be
t(p_t) \to \nu(p_1) + e^+(q_2) + b(q_f) + g(k), 
\label{eq4.1}
\ee
with $k^2 = \lambda^2$.
We again use the Low-Burnett-Kroll theorem 
\cite{Low:1958sn,Burnett:1967km,Fearing:1973eh}
shown in eq.~(\ref{eq2.22}) where for the   process 
in eq.~(\ref{eq4.1}) we have
\be
\begin{split}
  & J^\mu = J_t^\mu + J_f^\mu,\;\;\; J_t^\mu = \frac{2 p_t^\mu - k^\mu}{d_t},\;\;\;\; J_f^\mu = \frac{2 q_f^\mu + k^\mu}{d_f},
\end{split}
\label{eq:Jdec}
\ee
with $d_t = (p_t - k)^2 - \mt^2$ and $d_f = (q_f + k)^2$.

In order to factorise the phase space of the gluon
momentum from the rest of the decay phase space,
we employ a momentum mapping.
In variance with 
the case of the production sub-process, this mapping does not need to involve the top quark 
momentum and, hence, the production process 
remains unaffected.  Following 
ref.~\cite{Makarov:2023ttq}, we write  
\be
\begin{split}
  & q_f = p_f - k + \frac{ (p_{f}  k)}{(p_{2} p_f) } \;  p_{2},
  \;\;\;\; \qe = \left (1 - \frac{(\pfb k) }{(\pe  \pfb) } \right ) \pe.
\end{split}
\label{eq:mapdec}
\ee
Using this transformation, the phase space changes as follows~\cite{Makarov:2023ttq}
\be
   {\rm d}\Phi_D(p_t;p_1,\qe,q_{f},k) = {\rm d}\Phi_D(p_t;p_1,\pe,\pfb)\;[{\rm d} k]_\lambda  \left(1 + \frac{(\pe k)}{(\pfb \pe)} - \frac{(\pfb k)}{(\pfb  \pe)}
        +{\cal O}(\lambda^2) 
         \right ).
\ee

Since these momenta transformations do not impact $p_t$ and, therefore, the ``spin'' of the top quark as defined
by the production process,  the only addition to the unpolarised case for the width arises because
of the analog of the $S_t^\mu$ term in eq.~(\ref{eq2.22}) which is already ${\cal O}(\lambda)$ and, hence, can be
easily integrated over $k$.  We therefore find 
\begin{equation}
{\cal T}_\lambda  \left[{\rm d} \sigma^{R, \rm dec}_{PD} \right ]  = 2 \frac{ {\cal T}_\lambda  \left[ {\rm d} \Gammat^{R}(s_P) \right]  }{\Gammat} \; {\rm d } \sigma_{t} + {\cal T}_\lambda  \left[{\rm d} \sigma^{R, \rm dec}_{PD,\rm ~new} \right ],
\end{equation}
where in the first term formulas from unpolarised case can be employed except that  the leading order 
matrix element squared should be replaced 
with the polarised one. The second term is new;
after integration over the momentum of the soft gluon it evaluates to
\be
 {\cal T}_\lambda  \left[
    {\rm d} \sigma^{R, \rm dec}_{PD,\rm new} \right ] 
= -\frac{\alpha_s C_F}{2 \pi} \frac{\pi \lambda}{\mt} \;
2  \; \frac{ {\rm d} \sigma_{t}   }{\Gammat} \; s_{P,\mu} \;
\omega_{t f}^{\mu \nu} \frac{\partial}{\partial s_P^\nu} \; {\rm d} \Gammat(s_P).
\label{eq4.7}
\ee

We also need to compute virtual corrections and perform  mass and wave function  renormalisation  for the decay sub-process.   The virtual 
corrections are computed in the same way
as what was described for the production sub-process and in appendix B of 
ref.~\cite{Makarov:2023ttq}. We find 
\be
{\cal T}_\lambda 
\left [ {\rm d} \sigma_{PD}^{V, \rm dec} \right ]
= 2 \frac{{\cal T}_\lambda \left [ {\rm d} \Gammat^V(s_P) \right ]}{\Gammat}
\; {\rm d} \sigma_t 
+ {\cal T}_\lambda 
\left [ {\rm d} \sigma_{PD,\rm new}^{V, \rm dec} \right ],
\ee
where 
\be
{\cal T}_\lambda 
\left [ {\rm d} \sigma_{PD,\rm new}^{V, \rm dec} \right ] 
 = \frac{\alpha_s C_F}{2 \pi}
 \frac{\pi \lambda}{\mt} \;  2 \; \frac{{\rm d} \sigma_t}{\Gammat} \; s_{P,\mu} \; \omega^{\mu \nu}_{t f}
 \frac{\partial }{
 \partial s_P^\nu
 } \; {\rm d} \Gammat(s_P).
 \label{eq4.9}
\ee

The renormalisation  contributions are not 
affected by the polarisation of the top quark and, therefore, can be directly borrowed from the results in appendix B in ref.~\cite{Makarov:2023ttq} except 
that the differential decay width has to be computed for the polarised top quark. Hence, we write 
\be
{\cal T}_\lambda 
\left [ {\rm d} \sigma_{PD}^{\rm Ren, dec} \right ] 
= 2 \frac{{\cal T}_\lambda \left[{\rm d} 
\Gammat^{\rm Ren}(s_P)\right]}{\Gammat} \; {\rm d} \sigma_t.
\ee

\section{Redefinition of the top quark mass parameter}
\label{sect:mass}

In cases when the top quark production and the top quark decay are considered separately, it is known \cite{Makarov:2023ttq}
that the cancellation of ${\cal O}(\lambda)$ contributions is only possible if the production
cross section and the decay
rate are expressed through a short-distance top quark mass and \emph{not through the pole mass}. For reasons of technical convenience, 
 we performed the renormalisation in the on-shell scheme, similar to what was done in ref.~\cite{Makarov:2023ttq}.\footnote{A calculation that directly uses the mass parameter  defined in a short-distance scheme, see the appendix.
 } 
To derive the final results, we need to switch to a short-distance  mass parameter. We explain below 
how to do this in case when top quark
production and decay are considered simultaneously.

 As explained in  ref.~\cite{Makarov:2023ttq},
 we can  switch to a short-distance mass parameter
 by redefining momenta of final-state particles.  Our goal will be to do this  in such a way that,  when spin correlations are neglected, we obtain 
 formulas which  are identical to the 
 ones in ref.~\cite{Makarov:2023ttq}, 
 where production and decay are considered separately.

 We begin with the momenta
 transformations for particles that originate from top decay and write
\be
  q_f^\mu = \tpfb^\mu - \kappa q_t + \kappa\, \frac{(\tpfb q_t)}{(\tpfb \tpe)} \; \tpe^\mu,
  \;\;\;\; \qe^\mu = \tpe^\mu \left ( 1 - \kappa\, \frac{(\tpfb q_t)}{( \tpfb  \tpe)} \right ).
\label{eq5.1}
\ee
This momentum transformation leads to the following change in the decay phase space \cite{Makarov:2023ttq}
\be
   {\rm d}\Phi_D(q_t; q_f,\qe,q_1) = {\rm d}\Phi_D((1+\kappa ) q_t; \tpfb, \tpe,q_1)
   \left ( 1 + \kappa\, \frac{ ( \tpe q_t ) }{(\tpfb \tpe ) }
   - \kappa\, \frac{ (\tpfb q_t) }{(\tpfb  \tpe )}
   +{\cal O}(\lambda^2)
   \right ).
\ee
  It follows from the above equation that the mass of the decaying top quark becomes 
\be
{\barmt} = 
\sqrt{(1+\kappa)^2 q_t^2} = (1+\kappa) \mt.
\ee
Hence, 
\be
\barmt - \mt = \kappa \mt, 
\ee
which implies that $\kappa \mt $ is the shift in the mass parameter and where $\kappa$ is defined as in ref.~\cite{Makarov:2023ttq},
\begin{equation}
    \kappa = \frac{\alpha_s C_F}{2 \pi } \frac{\pi \lambda}{\mt}.
\end{equation}

To proceed further,  
it is convenient to  
define 
the top quark momentum that appears in the 
decay phase space 
\be
\tildep_t = (1+\kappa) q_t. 
\label{eq5.5}
\ee
When  the production and decay processes are considered together, the top quark momentum appears in the phase space of the production sub-process;
hence, the above redefinition will modify the production phase space and the matrix element. 
We begin with the analysis of the production 
phase space and write\footnote{To derive this formula, one needs to account for the fact that
  $q_t^2 \ne \mt^2$ a priori. Hence,
$\delta(q_t^2 - \mt^2) = \delta( (1 - 2\kappa) (\tilde{p}_t^2 - {\barmt}^2)) = (1 + 2 \kappa) \delta(\tilde{p}_t^2 - {\barmt}^2)$. }
\be
\begin{split} 
{\rm d}\Phi_P(...; q_t, q_d)
  &  = \frac{{\rm d}^4 q_t }{ (2 \pi)^3} \delta(q_t^2 - \mt^2) \; [{\rm d} q_d] \;(2\pi)^4 \delta(p_u + \pbi - q_t - q_d )
\\
& =  (1-2\kappa)  \frac{{\rm d}^4 \tildep_t}{(2\pi)^4}  \delta(\tildep_t^2 - \barmt^2)
  [{\rm d} q_d] (2\pi)^4 \delta(p_u + p_i - \tildep_t + \kappa \tildep_t  - q_d ).
\end{split} 
\ee
We then perform one more momentum redefinition but this time we 
change the  momenta in such a way that
the top quark remains on a (new) mass shell.
We write
\be
\begin{split} 
  & \tildep_t = p_t + \kappa  p_t - \frac{\kappa \mt^2}{ (p_t p_d) } p_d,\;\;\;\;\;
  q_d = \left ( 1  + \frac{\kappa \mt^2}{(p_t p_d)} \right ) p_d.
\label{eq5.6}
\end{split} 
\ee
This gives
\be
   {\rm d}\Phi_P(p_u,p_i; q_t, q_d)
   = {\rm d}\Phi_P(p_u,p_i; p_t , p_d)  \left (1 + \frac{\kappa \mt^2}{(p_t p_d)} 
   +{\cal O}(\lambda^2)
   \right ),
\ee
where $p_t^2 = {\barmt}^2$.

We note that the 
change  $\tildep_t \to p_t$ impacts the decay phase space \emph{again}. However, it is  easy to  solve this problem  because
this momentum change  can be written as a Lorentz transformation
\be
\tilde p_t^\mu = \Lambda_m^{\mu \nu} p_{t,\nu},\;\;\;\; 
\ee
where
\be
\Lambda_m^{\mu \nu} = g^{\mu \nu} + \kappa\, \omega^{\mu \nu}_{td}.
\label{eq5.10}
\ee
It follows that  the decay phase transforms as follows 
\be
   {\rm d}\Phi_D \left (\tilde p_t; \{ \tilde p_{\rm dec} \} \right )
   = {\rm d}\Phi_D( \Lambda_m  p_t,
   \{ \Lambda_m  p_{\rm dec} \} )
  = {\rm d}\Phi_D ( p_t,  \{p_{\rm dec}\} ).
\ee

We have worked out the momenta transformations required to modify the mass of a  top quark  in a process where it is
produced and then decays.   We now need to combine the several 
transformations and write down formulas  
that elucidate 
phase-space and 
matrix-element transformations 
of the full process. We use 
the representation
shown in eq.~(\ref{eq2.13})
where unpolarised decay 
width and polarised production cross section are  
combined.
We  use the  invariance of the decay 
matrix element squared $F_d$ under Lorentz transformations and find 
\begin{equation}
\begin{split} 
& {\rm d} \Phi_P(p_u,p_{i};q_t,q_d) F_{p}(p_u,\pbi;q_t,q_d,
s_D(q_t,q_{2}) )
  \; {\rm d} \Phi_D(q_t; q_{f}, q_2, q_{1}) F_d(q_t; q_{f}, q_{2}, q_{1})
  \\
  & \;
   = {\rm d} \Phi_P \, j_p \,  F_p{\Big  (...; \; p_t - \kappa \xi_d\, p_d , \;  \left(1+\kappa \xi_d \right) p_d 
, \; \Lambda_m s_D{\big(\left(1-\kappa\right) p_t, \left(1 - \kappa \xi_2 \right) p_{2} \big)} \Big )}
   \\ 
  &\;\;\;\;  \times
       {\rm d} \Phi_D \,
j_d \, F_d{\Big( \left(1-\kappa\right)p_t; \; \pfb + \kappa \delta \pfb, \; \left( 1-\kappa \xi_2 \right) \pe, \;   p_1 \Big)} + {\cal O}(\kappa^2),
\end{split}
\label{eq5.12}
\end{equation}
where  the phase spaces 
${\rm d} \Phi_{p,d}$ depend on the transformed 
momenta $\{ p\}$, 
$\Lambda_m$ is the 
boost defined 
in eq.~(\ref{eq5.10})
and 
\be
\begin{split} 
& j_p =  1 + \frac{ \kappa \mt^2}{(p_t p_d)},
\;\;\;\; j_d =   1 + \kappa\, \frac{(\pe p_t)}{(\pfb \tpe)}
   - \kappa\, \frac{(\pfb p_t)}{(\pfb  \pe)} ,
   \\
   & \xi_d =  \frac{ \mt^2}{(p_t p_d)}, 
   \;\;\;\; \xi_2 = \frac{(\pfb p_t)}{(\pfb \pe)},
   \;\;\;\;\delta \pfb^\mu =  -  p_t^\mu + \frac{(\pfb p_t)}{(\pfb \pe)} \; \pe^\mu.
   \end{split}
   \ee

To obtain ${\cal O}(\lambda)$ correction 
to the  cross section 
of the process in eq.~(\ref{eq2.1}) related to mass redefinition, 
we expand eq.~(\ref{eq5.12}) in $\kappa$ and keep 
linear terms.  These terms 
can be combined into three groups:
\begin{enumerate}
\item the term that originates from the expansion of $F_p$ caused by the ${\cal O}(\kappa)$ contribution to the  matrix  $\Lambda_{m}$ acting on $s_D$; 

\item all ${\cal O}(\kappa)$
terms that appear from the 
expansion of  the second line in eq.~(\ref{eq5.12})  
but without a term discussed in the previous item \emph{and} without 
a correction to the argument of the spin vector $s_D$;

\item terms that originate 
from the expansion of the third line 
in  
eq.~(\ref{eq5.12}) in $\kappa$ 
\emph{and} terms that originate from the
expansion of the argument of 
spin vector $s_D$ in function $F_p$.
\end{enumerate}

We now discuss these three groups of terms separately. 
The term in the first item is new. Terms in the second item provide the required 
contribution to cancel 
all ``old'' 
${\cal O}(\lambda)$ corrections to the production sub-process 
discussed in section~\ref{sect3}. Note that this cancellation occurs for the polarised matrix element squared $F_p(...,s_D)$ since this 
is what appears in eq.~(\ref{eq5.12}). 

The contribution of the third group of terms should, in principle, cancel all 
``old'' ${\cal O}(\lambda)$ terms to the decay sub-process,  described in section~\ref{sect4}. However, it follows from eq.~(\ref{eq5.12}) that 
this contribution lacks 
the polarisation vector  $s_P$,  which is present in  the similar contributions in  section~\ref{sect4}. Hence, to claim this cancellation, we need to put it back 
into the decay matrix element squared. 
This is possible because the following identity 
holds 
\begin{align}
  &
 F_p{\Big(...;p_t ,   p_d 
, s_D{\big( \left( 1-\kappa \right) p_t, \; \left(1 - \kappa \xi_2 \right) p_{2} \big)}
   \Big)}
F_d{\Big (\left(1-\kappa\right)p_t; \; \pfb + \kappa \delta \pfb, \; 
(1-\kappa \xi_2) \pe , \;   p_1 \Big )} 
\nonumber 
\\
& =
 F_p {\Big(...;p_t ,   p_d 
   \Big)}
F_d{\Big(\left(1-\kappa\right)p_t; \; \pfb + \kappa \delta \pfb, \; (1-\kappa \xi_2)\pe , \;   p_1, \; s_P{(p_t,p_d)} \Big)},
\end{align}
thanks to the relation between 
polarised production and decay cross sections 
shown in eq.~(\ref{eq2.19}).

We conclude that the only new term that we need to consider is the term in the first item 
that arises from the boost of the spin vector. It evaluates to 
  \be
   {\cal T}_\lambda \left [    {\rm d} \sigma^{\rm new}_{\rm mass}  \right ] 
         = -\frac{\alpha_s C_F}{2 \pi } \frac{\pi \lambda}{\mt} \; 2 \; \frac{ {\rm d} \Gammat}{ \Gammat}
         s_{D,\mu}\;  \omega_{td}^{\mu \nu}   \;
         \frac{\partial}{\partial s_{D}^{\nu}} \; {\rm d} \sigma_t(s_D).
         \label{eq5.15}
   \ee
Other terms that arise from the mass redefinition 
combine  with ``old'' contributions to the 
production and decay sub-processes and cancel 
in the same way  as  discussed in ref.~\cite{Makarov:2023ttq}.

\section{Results and corrections to observables }
\label{sect:kinematics}

The final result is obtained by combining 
the ${\cal O}(\lambda)$ contributions to single top production and decay process  derived 
in sections~\ref{sect3}, \ref{sect4} and \ref{sect:mass}. As we argued 
extensively during the calculation 
many   ${\cal O}(\lambda)$ contributions cancel in the sum; the only ones that survive involve  polarisation effects  which is an important new feature 
of a process with a long-lived particle that is first produced and then decays. 
They are obtained by adding 
eqs~(\ref{eq3.32}, \ref{eq3.41}, \ref{eq4.7}, \ref{eq4.9}, \ref{eq5.15}).
We find 
\begin{equation}
    {\cal T}_\lambda [ {\rm d} \sigma_{PD} ]
   = \frac{\alpha_s C_F}{2 \pi} \; \frac{\pi \lambda}{\mt} \; 2 \; \frac{{\rm d} \Gammat}{\Gammat} \;
    s_{D,\mu} \Big ( 2 \omega_{ti}^{\mu \nu} + 2 \omega_{dt}^{\mu \nu}
  + \frac{2 m_t^2 (\pbi p_d)}{(p_t \pbi)(p_t p_d) }  \omega_{i d}^{\mu \nu}  \Big  ) \frac{\partial }{\partial s_{D}^\nu }
  \; {\rm d} \sigma_t(s_D).
\label{eq6.1}
\end{equation}
 We can use eq.~(\ref{eq2.19}) as well as the relations   $s_P \cdot p_t = s_D \cdot p_t = 0$,  to find 
\begin{equation}
    {\cal T}_\lambda [ {\rm d} \sigma_{PD} ]
    = 
  -\frac{\alpha_s C_F}{2 \pi} \; \frac{\pi \lambda}{\mt} \; 
   \frac{{\rm d} \Gammat {\rm d} \sigma_t }{\Gammat} \;
   \frac{2 \mt^2 (\pbi p_d)}{(p_t \pbi)(p_t p_d) }  \;
   s_{D,\mu}  
   \omega_{i d}^{\mu \nu}  s_{P,\nu}. 
\label{eq6.3}
\end{equation}

The expression in eq.~(\ref{eq6.3}) assumes a particularly simple form in the 
\emph{top quark rest frame}.
Indeed, in this case 
\be
s_D^\mu = \left (0, \vec n_{2}
\right ) ,
\;\;\;
s_P^\mu = (0, \vec n_{d}
 ),
\ee
where $\vec n_{2}$ and $\vec n_d$ are unit vectors 
aligned with directions of the positron and $d$-quark in this frame, respectively.  This implies that 
\be
 \frac{2 \mt^2 (\pbi p_d)}{(p_t \pbi)(p_t p_d) }  \;
   s_{D,\mu}  
   \omega_{i d}^{\mu \nu}  s_{P,\nu}
=  2 \left ( 
(  \vec n_2 \cdot \vec n_i  )
- 
(  \vec n_2 \cdot \vec n_d  )
(  \vec n_i \cdot \vec n_d  )
\right ) 
= 2\, [\vec n_2 \times 
\vec n_d ] \cdot 
 [ \vec n_i \times \vec n_d],
 \label{eq:directionlinearterm}
 \ee
where $\vec n_i$ is the direction of the incoming $b$ quark in the top quark rest frame. 
We conclude that, in the top quark rest frame,  eq.~(\ref{eq6.3}) takes 
a remarkably simple form 
\be
    {\cal T}_\lambda [ {\rm d} \sigma_{PD} ]
    = -\frac{\alpha_s C_F}{2 \pi} \; \frac{\pi \lambda}{\mt} \; 
   \frac{{\rm d} \Gammat  {\rm d} \sigma_t }{\Gammat } \; 2\;
   [\vec n_{2} \times 
\vec n_d ] \cdot 
 [ \vec n_i \times \vec n_d]. 
 \label{eq6.6}
\ee

For the case, when the observable does not depend on decay momenta, we recover our previous result presented in  ref.~\cite{Makarov:2023ttq}. Indeed, integrating over the total phase space of the decay products and using the fact that 
\be
\int {\rm d} \Gamma_t \; \vec n_2 = 0,
\label{eq6.6a}
\ee
in the top quark rest frame, we  recover the 
stable-top-quark 
result~\cite{Makarov:2023ttq}
\be
    {\cal T}_\lambda [\sigma_{PD}]
    = 0.
\ee
Nevertheless, eq.~(\ref{eq6.6}) 
shows that, in general,  there \emph{is} an 
${\cal O}(\lambda)$ contribution to the differential cross section 
related to polarisation effects.   However, 
because we have used momenta redefinitions to derive this result, 
it is important to account for them also in the \emph{observables} 
since they  
are defined using the original momenta.

To this end, we consider an observable 
$X$ and study the following integral
\begin{equation}
O_X = \int {\rm d} \sigma_{PD} \; X.
\label{eq7.1}
\end{equation}
In principle, the  observable $X$ is  generic;  however,  we  would like to  focus  upon  observables that are used in practice to study  
polarisation effects in single top production \cite{CMS:2013rfa, Komm:2014fca, ATLAS:2022vym}. For this reason, we  assume that the observable 
$X$ depends on the top quark  momentum, the $d$-quark momentum, the incoming $b$-quark  momentum (the collision axis) and the positron momentum
\be
X = X(q_t,q_d,q_i,q_2).
\label{eq7.2}
\ee
When different contributions 
to eq.~(\ref{eq7.1}) are studied and different mappings are performed, there will be shifts in the arguments of the function $X$ that are proportional to the gluon momentum $k$
or to the mass-redefinition parameter $\kappa$. We are interested in terms that originate in the expansion of the function $X$ in these small parameters.

There are three contributions that affect the arguments of $X$:
real radiation in production, real radiation in decay and 
mass redefinition.  As the first step, we summarise the momenta redefinitions  for each of these 
contributions. Since none of these momenta redefinitions changes the collision axis, we will not show  $q_{i}$ among the arguments of $X$ in what follows. We find:
\begin{itemize}
\item Radiation in the production subprocess:
\begin{equation}
\begin{split}
    X \to  X \left(\Lambda p_{t} , \left( 1 - \frac{(\ptop k)}{(p_t p_d)} \right) p_d ,  \Lambda p_{2} \right),
\end{split}
\label{eq:Xprod}
\end{equation}
where 
\be
\Lambda^{\mu \nu}
= g^{\mu \nu} + 
\frac{p_d^\mu k^\nu - k^\mu p_d^\nu}{(p_t p_d)}.
\ee
\item Radiation in the decay subprocess:
\begin{equation}
\begin{split}
    X \to  X \left(p_{t}, p_d , \left( 1 - \frac{(p_{f} k)}{(p_{f} p_{2})} \right) p_{2} \right),
\end{split}
\label{eq:Xdec}
\end{equation}
\item Mass redefinition:
\begin{equation}
    X \to  X \left(\left( 1 - \kappa \right) \Lambda_m p_{t}, \left( 1 + \kappa \frac{\mt^2}{(p_t p_d)} \right) p_d ,  \left( 1 - \kappa \frac{(p_{f} p_t)}{(p_{f} p_{2})} \right) \Lambda_m p_{2} \right),
\label{eq:massred}
\end{equation}
where 
\be
\Lambda_m^{\mu \nu} 
= g^{\mu \nu} 
+ \kappa 
\frac{p_t^\mu p_d^\nu - p_d^\mu p_t^\nu}{(p_t p_d)}.
\ee
\end{itemize}

We then expand $X$ in series  for each 
of the three contributions and integrate 
over the gluon momentum $k$ where appropriate.
This is straightforward, and the matrix element squared is only needed in the eikonal approximation.   The result reads
 \begin{equation}
 \begin{split}
& \delta X = \frac{\alpha_s C_F}{2 \pi}\;
\frac{\pi \lambda}{m_t}
\Bigg [ 
\left ( p_t^\mu  - \frac{2 \mt^2}{(p_t p_{i})} p_{i}^\mu  \right )
\frac{ \partial X}{\partial p_{t}^\mu} 
-2 p_{2,\nu} 
 \left ( \omega_{t d}^{\nu \mu }  
 + \frac{\mt^2 (p_d p_{i})}{
 (p_t p_d) (p_t p_{i}) 
 } \omega_{d i}^{\nu \mu} 
 \right ) \frac{\partial X}{\partial p_2^{\mu}}
 \Bigg ].
\end{split}
\label{eq7.9}
\end{equation}
We note that the above result assumes   
that the mass parameter does not appear in the definition of 
the observable;  if this is not the case, 
the mass parameter needs to be replaced with 
$\sqrt{p_t^2}$.

Eq.~(\ref{eq7.9}) is applicable to any observable; the only constraint is  that 
it can only depend on the momenta of final-state 
particles shown in eq.~(\ref{eq7.2}).  Hence, 
we conclude that the complete linear correction to the expectation 
value of such an observable reads 
\be
\begin{split}
& {\cal T}_\lambda [{\cal O}_X]
 = \frac{\alpha_s C_F}{2 \pi} \; \frac{\pi \lambda}{\mt} \;  
 \int  \frac{{\rm d} \Gammat {\rm d} \sigma_t }{\Gammat} \;
 \Bigg  [  - 
   \frac{2 \mt^2 (p_i p_d)}{(p_t p_i)(p_t p_d) }  \;
   s_{D,\mu}  
   \omega_{i d}^{\mu \nu}  s_{P,\nu}
   \; X
\\
&  + \left(1-s_D \cdot s_P \right) 
\left  [ 
\left ( p_t^\mu  - \frac{2 \mt^2}{(p_t p_{i})} p_{i}^\mu  \right )
\frac{ \partial X}{\partial p_{t}^\mu} 
-2 p_{2,\nu} 
 \left ( \omega_{t d}^{\nu \mu }  
 + \frac{\mt^2 (p_d p_{i})}{
 (p_t p_d) (p_t p_{i}) 
 } \omega_{d i}^{\nu \mu} 
 \right ) \frac{\partial X}{\partial p_2^{\mu}} 
 \right ]
 \Bigg ].
   \end{split}
   \label{eq7.10}
\ee

We will now analyse this general formula.
First we note that  one can consider observables that depend on the 
top quark  momentum, 
but are inclusive with respect to  the momenta of its decay products. Then $X$ is a function of $p_t$ only.    For such observables, we can integrate 
over the momenta of the top quark decay products. Then, considering the integrand in the top quark rest frame 
and  using eq.~(\ref{eq6.6a}), we conclude that  the 
 first term on the right hand side 
 in eq.~(\ref{eq7.10})  vanishes. The second term then coincides with the correction to observable 
discussed in ref.~\cite{Makarov:2023ttq}
and the last term vanishes if  $X$ is a function of $p_t$ only.  
\\

There are also observables that are designed to study polarisation effects in single top production.   Perhaps the simplest observable  that belongs to this class is the 
one used by the CMS collaboration where 
the asymmetry between the direction of the outgoing light jet 
in single top production ($d$-jet in our case) 
and the direction of positron in top decay 
is studied in the top quark rest frame \cite{Komm:2014fca}. 
We can construct such an observable by simply multiplying  the production and decay
spin polarisation vectors, $s_P$ and $s_D$. 
Since in the top rest frame 
\be
s_D \cdot s_P = - \vec n_2 \cdot \vec n_d = - \cos \theta_{d2},
\ee
any function of this variable will provide 
a probe of  polarisation effects; the observable used by the CMS collaboration corresponds to 
\be
X_{\rm CMS} = \theta(-s_D s_P) - \theta(s_D s_P).
\ee
Using eq.~(\ref{eq7.9}), 
it is easy to show that 
\be
\delta X_{\rm CMS}
 = 0,
\ee
which then implies that the
only relevant term in eq.~(\ref{eq7.10}) 
that contributes for such observables is the 
first term in the integrand of eq.~(\ref{eq7.10}).  

It is interesting to note  that one can arrive at the same result without any  computation. 
In fact, there is a 
simple argument that can be used 
to argue that 
for any  observable $X$ that depends
upon the \emph{directions} of $p_t$, $p_d$ and $p_2$ \emph{only}\footnote{We note that   $s_D$ and $s_P$ belong to this category.} there cannot be any change in $X$ after the  remapping described in this paper.  To illustrate this argument, consider radiation in the production. According to  eq.~(\ref{eq:Xprod})
the momenta redefinitions lead to
\begin{equation}
    X(p_t, p_d, p_2 )\rightarrow
    X\left(\Lambda p_t, \left(1-\frac{(p_t k)}{(p_t p_d)}\right)p_d, \Lambda p_{2}\right) 
    = X\left(\Lambda p_t, \left(1+\frac{(p_d k)}{(p_t p_d)}\right)p_d, \Lambda p_{2}\right),
\end{equation}
where in the last step we used the 
fact that the observable  $X$ depends on the direction of 
$p_d$. This implies that  the exact form of  rescaling  
is irrelevant, and we can change it at will.  Since 
\be
\left(1+\frac{(p_d k)}{(p_t p_d)}\right) p_d = \Lambda p_d,  
\ee
we find 
\be
X\left(\Lambda p_t, \left(1+\frac{(p_d k)}{(p_t p_d)}\right)p_d, \Lambda p_2 \right) 
= X\left(\Lambda p_t,  \Lambda p_d, \Lambda p_2\right) 
= X\left( p_t,   p_d,  p_2\right),
\ee
where in the last step  Lorentz invariance 
of the observable was used.  Hence, we conclude 
that the momenta redefinitions employed in the 
description of the real emission in  production do not change an observable 
which depends on directions of final-state 
particles. The same reason also applies to the momenta transformations employed to describe 
radiation in decay and the  mass redefinition. 

To complete the analysis of the CMS asymmetry,  we need to understand 
the fate of the first term in 
eq.~(\ref{eq7.10}).  Considering  
this term in the top rest frame, 
we find that it 
involves the following integral 
\be
\int {\rm d} \Gammat  \; [ \vec n_{2} \times 
\vec n_d ] \;  X(\vec n_2 \cdot \vec n_d ).
\label{eq7.17}
\ee
Since for any function $X$
\be
\int {\rm d} \Gamma_t \; \vec n_{2}\;
 X(\vec n_{2} \cdot \vec n_d )
\sim \vec n_d ,
\label{eq14.24}
\ee
the integral in eq.~(\ref{eq7.17}) vanishes.  We conclude that the asymmetries  in single top production studied by the CMS collaboration \cite{CMS:2013rfa,Komm:2014fca} are not affected by the non-perturbative effects 
that can be modelled with renormalons. 
\\

A more complex polarisation observable was studied by the ATLAS collaboration \cite{ATLAS:2022vym}. To define it, a reference system in the top rest frame is 
introduced, where 
the three  axes are\footnote{We note that in our calculation $\vec n_i$ denotes the direction of the incoming $b$-quark in the top quark rest frame, whereas in the ATLAS paper \cite{ATLAS:2022vym}
the direction of the incoming light quark is chosen to define the reference system.   These vectors are not back-to-back in the  top quark  rest frame 
but, thanks to momentum conservation, in this reference frame their 
vector products with 
$\vec n_d$ are  the same up to a sign.} 
\be
\vec e_z = \vec n_d, \;\;\;\;
\vec e_y = \frac{\vec n_i \times \vec n_d}{
| \vec n_i \times \vec n_d|}, 
\;\;\; \vec e_x = \vec e_y \times \vec e_z 
 = \frac{\vec n_i \times \vec n_d}{|\vec n_i \times \vec n_d| } \times \vec n_d. 
 \label{eq14.25}
\ee

The observable $Q$ is defined as follows  
\begin{equation}
Q(\vec n_2, \{ \vec e\}) = 4 \theta(\vec n_2 \cdot \vec e_z) 
+ 2 \theta(\vec n_2 \cdot \vec e_x ) 
+ \theta( \vec n_2 \cdot \vec e_y ).
\end{equation}
We now determine the expectation value  of $Q$ at leading order and the non-perturbative correction to it. First, writing the leading order cross section using the reference frame described above, 
we obtain 
\be
{\rm d} \sigma_{PD} = {\rm d}\sigma_t \frac{{\rm d} \Gammat}{\Gammat}
\left ( 1 + \vec e_z \cdot \vec n_2 \right ).
\label{eq14.28}
\ee
If we integrate over the top quark decay products without imposing any cuts 
on final-state particles, the following equations hold
\be
\begin{split}
& \int {\rm d} \Gammat \;
 \theta(\vec n_{2} \cdot \vec a) 
= \frac{1}{2} \Gammat,
\;\;\;\;\; \int {\rm d} \Gamma_t \;
 \vec n_{2} \; \theta(\vec n_{2} \cdot \vec a) 
=  \frac{1}{4} \Gamma_t \; \vec a,
\label{eq14.29}
\end{split}
\ee
where $\vec a$ is an arbitrary unit vector. We use 
eq.~(\ref{eq14.29})
together with the leading order cross section in eq.~(\ref{eq14.28}) to find 
\be
\langle Q(\vec n_2, \{\vec e\}) \rangle 
= \frac{\int {\rm d} \sigma_{PD} \;  Q(\vec n_2, \{\vec e\}) } {
\int {\rm d} \sigma_{PD} \;  } =\frac{9}{2}.
\ee

To compute the power corrections to this result, we need to combine 
the corrections to the cross section 
and to the observable.   We begin with the latter.  The correction 
to the observable is computed using 
eq.~(\ref{eq7.9}). To apply 
this equation to the observable 
$Q$, we should write it in a Lorentz-covariant form. 
To this end, we write 
\begin{equation}
Q(\vec n_2, \{ \vec e\}) = 4 \theta(\hat Q_z) 
+ 2 \theta(\hat Q_x) 
+ \theta(\hat Q_y),
\end{equation}
with
\begin{equation}
\begin{split}
    \hat Q_x = \vec n_{2} \cdot \vec e_x =& \, \frac{1}{| \vec n_i \times \vec n_d|} \; \big[\left(\vec n_{2} \cdot \vec n_d\right) \left(\vec n_i \cdot \vec n_d\right) - \left(\vec n_i \cdot \vec n_{2}\right) \big],
    \\
    \hat Q_y = \vec n_{2} \cdot \vec e_y =& \, \frac{1}{| \vec n_i \times \vec n_d|} \frac{p_t^2}{\left(p_t  p_{2}\right) \left(p_t  p_{i}\right) \left(p_t  p_d \right)} \;\epsilon_{\mu \nu \rho \sigma}\; p_t^\mu p_{2}^\nu p_{i}^\rho p_d^\sigma,
    \\
    \hat Q_z = \vec n_{2} \cdot \vec e_z =& \, \vec n_{2} \cdot \vec n_d,
\end{split}
\end{equation}
and note that the covariant  generalisation of the scalar product of two vectors in the top rest frame is given by   
\begin{equation}
    \vec n_i \cdot \vec n_j = 1 - \frac{p_t^2 \left(p_i  p_j\right)}{\left(p_t p_i\right)\left(p_t  p_j\right)}.
\end{equation}
After that, the calculation becomes straightforward. We obtain 
\begin{equation}
\begin{split}
    \delta \hat Q_x = \, 0,   \qquad\qquad\qquad    \delta \hat Q_y = \, 0,  \qquad\qquad\qquad   \delta \hat Q_z = \, 0,
\end{split}
\label{eq:shiftobservableATLAS}
\end{equation}
so that also in this case there is no change in the observable
\begin{equation}
    \delta Q = 0.
\end{equation}

Written in the reference system defined in 
eq.~(\ref{eq14.25}), the  non-perturbative shift in the cross section  shown in eq.~(\ref{eq6.6}) becomes
\be
    {\cal T}_\lambda [ {\rm d} \sigma_{PD} ]
    = \frac{\alpha_s C_F}{2 \pi} \; \frac{\pi \lambda}{m_t} \; 
   \frac{{\rm d} \Gamma_t {\rm d} \sigma_t }{\Gamma_t} \; 2 \; |\vec n_i \times \vec n_d|\;
   ( \vec e_x \cdot \vec n_2 ). \; 
 \label{eq14.26}
\ee
We then integrate the product of this quantity  with the observable  $Q$ over the 
top quark decay products, and find
\be
\begin{split}
    & \int 
    {\cal T}_\lambda [ {\rm d} \sigma_{PD} ]
    \; Q(\vec n_{2}, \{ \vec e\}) \; 
    = \frac{\alpha_s C_F}{2 \pi} \; \frac{\pi \lambda}{\mt} \; 
    {\rm d} \sigma_t \frac{2
    |\vec n_i \times \vec n_d| }
   {\Gammat} \;\;  \vec e_x \cdot 
    \int {\rm d} \Gammat 
    \;  Q (\vec n_2, \{ \vec e\})\; 
  \vec n_2
  \\
&  = 
  \frac{\alpha_s C_F}{2 \pi} \; \frac{\pi \lambda}{\mt} \; 
    {\rm d} \sigma_t \frac{4 
     |\vec n_i \times \vec n_d| 
    }{\Gammat} \;  \vec e_x \cdot 
    \int {\rm d} \Gammat 
    \;  \theta(\vec  n_{2} \cdot \vec e_x)\; 
  \vec n_{2}
  = 
   \frac{\alpha_s C_F}{2 \pi} \; \frac{\pi \lambda}{\mt} \; 
    {\rm d} \sigma_t
    \;|\vec n_i \times \vec n_d|. 
  \end{split}
   \ee
We  use this equation to determine the non-perturbative correction to the expectation value of the  
observable  $Q$
\be
\langle Q \rangle = \frac{1}{\sigma_t} \int{\rm d} \sigma_{t} 
\left ( \frac{9}{2}  + \frac{\alpha_s C_F}{2 \pi} \; \frac{\pi \lambda}{\mt} \; | \vec n_i \times \vec n_d | 
 \right ),
\ee
at fixed center-of-mass collision energy $\sqrt{s}$.
We note that in the center-of-mass frame of partonic collision, the absolute value of the vector product of $\vec n_i $ and $\vec n_d$ reads 
\be
| \vec n_i \times \vec n_d|
=  \sqrt{\frac{4 \mt^2\; s\, t\, u}{\left(s-\mt^2\right)^2 \left(\mt^2-t\right)^2}}
= 
\frac{ 
2 \mt\, s\, {p_d}_{\perp}
}{(\mt^2-t) (s-\mt^2)},
\ee
where ${p_d}_{\perp} $ is the transverse momentum of the $d$-jet relative to the collision axis. We note that in the last step  we used the fact that $t\,u=s\, {p_d}^2_{\perp}$. 

Integrating over the scattering angle, we 
find 
\begin{equation}
\frac{1}{\sigma_t} \int {\rm d} \sigma_t \; | \vec n_i \times \vec n_d  |
=  f_Q(s,\mt,m_W),
\end{equation}
where  the function $f_Q$ reads 
\begin{equation}
f_Q(s,\mt,m_W) = 
\frac{\pi  \mt  m_W \sqrt{s} \sqrt{\bar s} \left(-\mt^4+\mt^2 \left(m_W^2+s\right)-2 \mt m_W \sqrt{s} \sqrt{\bar s} + m_W^2 s\right)}{\left(\mt^2-m_W^2\right)^2 \left(s-\mt^2\right)^2},
\end{equation}
and we have defined the quantity
\begin{equation}
    \bar s = s - \mt^2 + m_W^2.
\end{equation}
The non-perturbative correction to the 
expectation value of the variable $Q$ 
in proton collisions is obtained by convoluting 
the above result with 
parton distribution functions.  In principle,  since  the function $f_Q(s,\mt,m_W)$ depends on 
the center-of-mass energy,  parton distribution functions do not decouple. However, in practice, 
$f(s,\mt,m_W)$ is a slowly changing 
function of $s$. Indeed, it changes from   the value 
\begin{equation}
\lim_{s \to \mt^2} f(s,\mt,m_W) = \frac{\pi}{4} \approx 0.785,
\end{equation}
at the threshold, to 
\begin{equation}
\lim_{s \to \infty} f(s,\mt,m_W) =   \frac{ \pi\, \mt\, m_W}{(\mt+m_W)^2}
\approx 0.68,
 \end{equation}
at $s = \infty$ for physical values of $m_t$ and $m_W$. Hence, we find the following estimate  for  the ATLAS variable $Q$ 
in proton-proton collisions 
\be
\langle Q \rangle 
\approx \frac{9}{2}  + \frac{\alpha_s C_F }{2 \pi } \; \frac{\pi \lambda}{\mt}
\; \frac{\pi}{4}.
\ee

The above result does not account for realistic event selection 
criteria which in many ways introduce additional directions into the 
integration over top quark decay products. However, it does illustrate 
the point that non-perturbative effects that we discuss in this paper have a small but direct impact on the measured values of the top quark polarisation observables  at hadron colliders. 

An outstanding problem in collider physics is the measurement of the top quark mass with an 
ultrahigh precision in a credible way \cite{Azzi:2019yne}.
The tricky issue is the control (or lack of it) of non-perturbative corrections,  which 
is very hard to do for exclusive observables that are used currently for the highest-precision measurements.  In this regard, suggestions were made to study lepton observables from top quark decay because they are considered 
to be less prone to contaminations by non-perturbative effects.   Interestingly, our analysis allows us to make exact statements to this effect, albeit in the narrow width approximation.

To this end, consider the following quantity 
\be
L_\perp = |\vec p_2 \cdot \vec e_y |, 
\ee
where $\vec p_2$ is the positron  
three-momentum in the top quark rest frame. The vector $\vec e_y$
is defined in eq.~(\ref{eq14.25}); it is orthogonal to the collision plane 
of single top production process. 
Hence, $L_\perp$ measures  the component of the lepton momentum that points \emph{outside} 
of the collision plane.
We are interested in computing the average value  of $L_\perp$.  Since 
\be
\int {\rm d} \Gamma_t \; \vec n_{2} \; \theta( \vec p_{2}  \cdot \vec e_y) \sim \vec e_y, 
\ee
it follows that neither the term $s_P \cdot s_D $ in the leading order cross section, nor 
the power correction ${\cal T}_\lambda [ {\rm d} \sigma_{PD}] $ receive contribution from the above integral.  
Since it  is also easy to check that $L_\perp$ does not 
receive any corrections from momenta redefinitions,  $\delta L_{\perp} = 0$, 
it follows that 
\be
\langle L_\perp \rangle 
 =  \frac{1}{\Gammat}
 \int {\rm d} \Gammat \; L_\perp  
 + {\cal O}(\lambda^2) 
  = \frac{1}{2 \Gammat}
   \int {\rm d} \Gammat \; \frac{ ( p_{2} p_t )}{\mt}
  = \frac{\mt^2 + m_W^2}{8 \mt} + {\cal O}(\lambda^2),
\ee
where in the last step we  employed the  narrow width approximation for the $W$-boson to integrate over 
the positron  momentum.

In summary,  the point of the above calculation is to  demonstrate  that there are no linear power corrections 
to $L_\perp$,
so that the (short-distance) top quark mass can be determined  from this observable with very high precision.
An obvious reservation is that the above calculation is  valid in the case when no fiducial cuts are  imposed on  final-state particles 
but it is also obvious that to perform such 
measurements in practice,  cutting-edge simulations are required that account for 
perturbative and parton shower effects.

\section{Conclusions}
\label{sect:concl}
In this paper we have studied linear power corrections to the process of single top production followed by the top quark decay.  Our primary interest is  the impact of  top quark instability on these corrections. 
Working in the narrow width approximation, we have found that linear power corrections do affect the top quark production cross section if the top quark  is allowed to decay, at  variance with 
the case of a stable top quark that was studied earlier in ref.~\cite{Makarov:2023ttq}.

The non-perturbative corrections that we have found in this article do affect measurements of the top quark polarisation in such processes, and also influence the  kinematic distributions 
of leptons in top quark decays that were suggested  as ``clean'' observables for measuring the top quark mass.  However, the  particular form of  power corrections, that we derived in this paper, allows us to show that ``out of the collision plane'' component of the positron momentum from top quark decays does not receive linear non-perturbative corrections.  Since 
the average value of this observable depends on $m_t$,  it  is  an  interesting candidate for measuring the top quark mass. 

Finally, the  results discussed in this paper are obtained in the narrow width approximation for the top quark which corresponds to an unphysical limit  $\Gamma_t \ll \Lambda_{\rm QCD}$.  The next important step is  to extend these result to the \emph{physical} case 
$\Gamma_t \gg  \Lambda_{\rm QCD}$. Then, the analysis becomes significantly more complicated because top quark production process and top quark decay do not factorise any more. Nevertheless, we  hope that our understanding of non-perturbative power corrections 
to top quark production processes achieved in 
this paper as well as 
in refs~\cite{Makarov:2023ttq, Makarov:2023uet} will allow 
us to successfully  analyse this challenging problem. 

\section*{Acknowledgments}
The research of K.M. and S.M. was supported by the German Research Foundation (DFG, Deutsche Forschungsgemeinschaft) under grant 396021762-TRR 257.
P.N acknowledges support by the Humboldt Foundation. M.A.O. was supported by the ANR PIA funding under grant ANR-20-IDEES-0002.

\appendix

\section{Alternative derivation of power corrections}
\label{sec:appendix}

The goal of this appendix is to discuss 
an alternative derivation of power corrections to the single top production 
and decay process. 
Here we  deal directly 
with the amplitudes as opposed to amplitudes 
squared, as was done 
in earlier papers \cite{Makarov:2023ttq,Makarov:2023uet}. Below, we first discuss  the Born amplitude and cross section and then continue with the real emission and virtual corrections to the  top quark  production and decay process.  
We use directly  a short-distance mass scheme for the top quark mass in the calculation and 
we  explain  how to do this by considering 
the self-energy insertion 
in  the (nearly) on-shell  top quark line.

\subsection*{The Born cross section}
The Born diagram for single top production and decay is
shown  in Fig~\ref{fig:Born}.
The colour structure of  this process is quite simple. We will ignore it for now and reconstruct it at the end. 
In the narrow width approximation, we write the Born amplitude as
\begin{equation} \mathcal{B}_{P D} = \frac{i}{p_t^2 - \mt^2 + i \mt \Gammat}  \bar{u}
   (p_f) N_D  \left[ \sum_{\lambda_s = \pm 1} u (p_t, s, \lambda_s)  \bar{u}
     (p_t, s,\lambda_s) \right] N_P u (p_i), 
     \label{eqb1}
\end{equation}
where the subscripts  $P$ and $D$ denote production and decay,
and we have displayed explicitly  the   $b_i - t - b_f$ fermion line. 
The functions $N_P$ ($N_D$)
contain all remaining structures pertinent to the production and decay processes.  We assume that 
a quantisation axis $s$, satisfying the conditions $s^2=-1$ and $p_t\cdot s=0$, has been chosen  
for the top quark spin.  We denote  
the signs of the
top quark spin along
the quantisation axis $s$ 
with $\lambda_s$. We then define 
\begin{equation}
\begin{split} 
  \mathcal{B}_P (s,\lambda_s) & =  \bar{u} (p_t, s,\lambda_s) N_P u (p_i),
  \\
  \mathcal{B}_D (s,\lambda_s) & =  \bar{u} (p_f) N_D u (p_t, s, \lambda_s),
  \end{split} 
\end{equation}
and write the Born cross section for single top production and decay  as follows
\begin{equation}
\begin{split}
  | A |^2 = & \frac{1}{2 \mt \Gamma_t} 2 \pi \delta (p_t^2 - \mt^2)  |
  \mathcal{B}_{P D} |^2,
  \\
  | \mathcal{B}_{P D} |^2 = & \sum_{\lambda_s, \lambda'_{s}} [\mathcal{B}_P (s,\lambda_s) 
   \mathcal{B}^*_P (s,\lambda'_{s})]  [\mathcal{B}_D (s,\lambda_s)  \mathcal{B}_D^* (s,\lambda'_{s})],
\label{eq:MPDSQ}
\end{split}
\end{equation}
where we have used the narrow width approximation, see  eq.~(\ref{eq:narrowwidth}).
The form of the amplitude 
in eq.~(\ref{eq:MPDSQ})
is the expected
product of spin correlation matrices. In single top production process, a further simplification
occurs since  there are  choices of  the 
top quark spin quantisation axes such as 
 $\mathcal{B}_D (s,-1) = 0$ or  
$\mathcal{B}_P (s,-1) = 0$.  The fact that such  quantisation axes must exist  is a consequence of the fact that the helicities of all massless particles in single top production and decay are fixed by the charged-current
interactions, so that also the top quark must be in a pure spin 
state.\footnote{This suggests that this
property should also be valid in a class of single top production processes with the
addition of colour-neutral particles with definite spin.} If we call such a  quantisation axis for the production process $s_P$,
we can  write 
\begin{equation}
\begin{split}
  \mathcal{B}_P(s,\lambda_s)= \bar{u}(p_t,s,\lambda_s) B =&\bar{u}(p_t,s,\lambda_s) \left[\frac{1+\gamma_5{\slashed s}_P}{2}
    +\frac{1-\gamma_5{\slashed s}_P}{2}\right] B
    \\
  =&\bar{u}(p_t,s,\lambda_s) \frac{1+\gamma_5{\slashed s}_P}{2}B.
  \label{eq.a4a}
\end{split}
\end{equation}
In eq.~(\ref{eq.a4a})   $B$ denotes whatever is left of $\mathcal{B}_P(s,\lambda_s)$ when the $\bar{u}$ spinor is removed, and
the last step follows from the fact that 
\begin{equation}
\bar{u}(p_t,s,\lambda_s) \left[\frac{1-\gamma_5{\slashed s}_P}{2}
\right ], 
\end{equation}
is an eigenstate of the projection of the top quark  
spin operator  on the 
axis $s_P$ with an eigenstate $-1/2$, which 
by assumption does not contribute to the production process. 

Upon squaring the amplitude, we find 
\begin{equation}
\begin{split}
  \left|\mathcal{B}_P(s,\lambda_s)\right|^2 =& \bar{B} \frac{1+ \gamma_5{\slashed s}_P}{2} u(p_t,s,\lambda_s)\bar{u}(p_t,s,\lambda_s)
  \frac{1+ \gamma_5{\slashed s}_P}{2}B
  \\
=&
  \bar{B} \frac{1+\gamma_5{\slashed s}_P}{2}\left[ ({\slashed p}_t+\mt)
  \frac{1+ \lambda_s \gamma_5{\slashed s}}{2} \right] \frac{1+\gamma_5{\slashed s}_P}{2} B.
  \label{eqa7}
  \end{split}
\end{equation}
Working out the simple Dirac algebra we find
\begin{equation}
  \frac{1+\gamma_5{\slashed s}_P}{2}\left[({\slashed p}_t+\mt)
  \frac{1+ \lambda_s \gamma_5{\slashed s}}{2}\right]  \frac{1+  \gamma_5{\slashed s}_P}{2} =
  \frac{1-\lambda_s s\cdot s_P}{2}  ({\slashed p}_t+\mt)  \frac{1+\gamma_5{\slashed s}_P}{2}.
\end{equation}
We then insert this result 
into eq.~(\ref{eqa7})  and obtain 
\begin{equation}
\left|\mathcal{B}_P(s,\lambda_s)\right|^2=\frac{1- \lambda_s s\cdot s_P}{2} \sum_{\pm \lambda'_s}\mathcal{B}_P^*(s_P,\lambda'_s) \mathcal{B}_P(s_P, \lambda'_s)
  = \frac{1-\lambda_s s\cdot s_P}{2} \left|\mathcal{B}_P\right|^2 ,
  \label{eq:MPSSQ}
\end{equation}
where we have introduced the notation $|\mathcal{B}_P|^2=\sum_{\lambda_s} |\mathcal{B}_P(s,\lambda_s)|^2$. 

A similar formula  
can be derived for the 
decay amplitude, 
\begin{equation}
  \left|\mathcal{B}_D(s,\lambda_s)\right|^2= \frac{1 -  \lambda_s s \cdot s_D}{2} \left|\mathcal{B}_D\right|^2 , \label{eq:MDSSQ}
\end{equation}
where the proper quantisation axis $s_D$ differs  from the one in the production. 
Combining the results for 
the production and decay amplitudes, we obtain  
\begin{equation}
  | \mathcal{B}_{P D} |^2 = \frac{1-s_P\cdot s_D}{2} \left|\mathcal{B}_P\right|^2 \left|\mathcal{B}_D\right|^2.
\label{eq:bpdsq}
\end{equation}
This result can be derived from eq.~(\ref{eq:MPDSQ}) by choosing the quantisation axis to be either  $s_P$ or $s_D$ in which case only a single term $\lambda_s = 1$ contributes to sums over spin projections,
and using  either eq.~(\ref{eq:MPSSQ}) or~(\ref{eq:MDSSQ}). 
We also note  that the differential decay 
width takes the form
\begin{equation}
\mathd \Gammat = \frac{1}{4\mt} \mathd \Phi_{D} |\mathcal{B}_D|^2,
\end{equation}
where we had to divide by two for the
spin average. On the other hand
our expression for the differential
cross section (ignoring spin and colour averages for the initial fermions) is given
by
\begin{equation}
\mathd \sigma_{PD}=\frac{1}{2 \mt \Gammat}
\mathd\Phi_P\mathd \Phi_D|\mathcal{B}_D|^2
|\mathcal{B}_P(s_D)|^2
= 2 \frac{\mathd \Gammat}{\Gammat}
\mathd \sigma(s_D),
\end{equation}
which agrees with eq.~(\ref{eq2.13}).
For the case of single top production depicted in Fig.~\ref{fig:Born}, we can easily  identify the quantisation axes 
$s_P$ and $s_D$. 
 We begin by computing $s_D$.  The decay amplitude is proportional to
 \begin{equation}
 \begin{split}
\mathcal{B}_D \sim& \bar u_f \gamma^\mu (1-\gamma_5)
u_t \; {\bar u}_1 \gamma_\mu (1-\gamma_5)  v_2
= -\bar u_f \gamma^\mu (1-\gamma_5)
u_t \; {\bar u}_{2R} \gamma_\mu (1+\gamma_5)  v_{1R}
\\
=&-  \left [ {\bar u}_{2R} (1-\gamma_5) u_t \right ] \; 
\left [ {\bar u}_f (1+\gamma_5) v_{1R}
\right ],
\end{split}
\end{equation}
where we have introduced the charge conjugate spinors $u_{2R}$ and $v_{1R}$
for the positron and the neutrino, and 
the last step uses a  Fierz  identity. The subscript  $R$ on the conjugate spinors is to remind that they are right-handed, i.e. 
$\bar u_{2,R}  \gamma_5 = - \bar u_{2,R}$.
We write
\begin{equation}
\begin{split}
  \bar u_{2,R}(p_2) u (p_t, s) = &   \frac{1}{2}
  {\bar u}_{2,R} (p_2) \left( 2 + \gamma_5 \frac{\mt}{(p_2 p_t)} \left(
  {\slashed p}_2 - \frac{(p_2 p_t)}{\mt^2} ( {\slashed p}_t - \mt ) \right)
  \right) u (p_t, s)  \\
  = & \frac{1}{2} 
  \overline{u}_{2,R} (p_2)
 \left( 1 + \gamma_5 \frac{\mt}{(p_2
   p_t)} \left( {\slashed p}_2 - \frac{(p_2 p_t)}{\mt^2} {\slashed p}_t \right) \right)
  u (p_t, s)  \\
   = & \overline{u}_{2,R} (p_2) \frac{1 + \gamma_5 {\slashed s}_D}{2} u (p_t, s),
\label{eq:spinproj} 
  \end{split}
\end{equation}
where $s_D$ reads 
\begin{equation}
  s_D^\mu = \frac{\mt}{(p_2 p_t)}\, p_2^\mu - \frac{1}{\mt}\, p_t^\mu,
\label{eq:sD}
\end{equation}
and satisfies the  conditions $s_D^2 = - 1$ and $s_D \cdot p_t = 0$.   We note that in deriving 
eq.~(\ref{eq:spinproj}), 
we used 
the fact that the spinors $\bar u_{2,R}$ and $u(p_t,s)$ 
satisfy the respective Dirac equations, and that 
$\bar u_{2,R}  \gamma_5 = - \bar u_{2,R}$
as follows from its definition. It follows 
from eq.~(\ref{eq:spinproj}) that 
the top quark in the decay is  polarised along 
the axis $s_D$,  and we will refer 
to this quantity as the top quark spin vector in the decay. 

Repeating the same calculation for the production amplitude, we easily find that top quarks are produced 
polarised along the quantisation axis which is  given by the following equation 
\begin{equation} s_P^\mu = \frac{\mt}{(p_d p_t)}\, p_d^\mu - \frac{1}{\mt}\, p_t^\mu. 
\label{eq:sP}
\end{equation}
Again, we will refer to this vector as the top 
quark spin vector in the production. 
Furthermore, in the following we will use a simplified notation, where omitting the $\lambda_s$ argument implies that it is 
taken equal to one.

\subsection*{Real corrections in production}

We will use the letter $q$ rather than $p$ to indicate momenta 
of particles that are affected by recoil when
a soft gluon is emitted. We will also denote the top spin vector as $s_q$, since it must be
orthogonal to $q_t$.
Momentum conservation is  given by
\begin{equation}
  p_i + p_u= q_t  + q_d +k.
\end{equation}
The difference between $q$'s and $p$'s (and between $s$ and $s_q$) are of order $k$, so we can change $q$'s into $p$'s
and $s_q$ into $s$ when dealing with subleading terms.

It was mentioned several times that to compute linear power corrections, we only need to consider gluon radiation off the heavy quark line. We split this  contribution into two diagrams, one that describes radiation off  the final-state top quark,  and the 
other one that describes gluon 
radiation off the $b$ quark 
in the initial state. 

We begin by computing  the  contribution to the amplitude of the gluon emission from the final state top quark. It is given by 
\begin{equation}
\begin{split}
\mathcal{R}^\mu_{P,{\rm f}} = & \bar{u} (q_t, s_q) \gamma^\mu \frac{{\slashed q}_t +
  {\slashed k} + \mt}{(q_t + k)^2 - \mt^2} N_P (q_t + k,q_i) u (q_i) \\
  = &   \bar{u} (q_t, s_q) \; \frac{   2 q_t^{\mu} + k^{\mu}   +   \sigma^{\mu \nu}
  k_{\nu}}{(q_t + k)^2 - \mt^2} 
  \; 
  N_P (q_t + k,q_i)
  u (q_i),\label{eq:RPfmu}
\end{split}
\end{equation}
where the gluon polarisation vector has been 
omitted. 
For simplicity, we have omitted the arguments of $\mathcal{R}_{P,\rm f}$. We should remind
the reader, however, that it depends upon all the $q$ and $p$ momenta, and upon the spin vector $s_q$. The arguments in  $N_P$ show that this function depends on  the $q$ momenta.
$N_P$ is similar to the  Born diagram case, except that the incoming top quark momentum is off-shell. Nevertheless, it is a 
well-defined function of the  external momenta.
We are interested in the leading  ${\cal O}(k^{-1})$ and next-to-leading ${\cal O}(k^0)$
terms in the limit of small gluon momentum $k$. 
When performing the manipulations below, we will always 
discard
terms that vanish  in the $k \to 0 $ limit.
\\

We focus on  the term $\sigma^{\mu\nu}k_\nu$ in  eq.~(\ref{eq:RPfmu})
acting on the $\bar{u}$ spinor. Consider the following equation
\begin{equation} \bar{u} (q_t, s_q) (1+a_\mu\sigma^{\mu \nu} k_{\nu}) =
  \bar{u}_a,
\end{equation}
where $u_a$ is defined as 
\begin{equation}
u_a \equiv (1+k_{\nu} \sigma^{\nu \mu} a_\mu) u (q_t, s_q),
\end{equation}
and $a^\mu$ is an arbitrary four-vector.
We note that, up to an irrelevant phase, 
a general Lorentz transformation of a spinor is given by  the following 
expression 
\begin{equation}
    \hat{S}(\Lambda) u(p,s) 
    = u(\Lambda p, \Lambda s),
\end{equation}
where for  an infinitesimal transformation 
\begin{equation}
\Lambda_{\alpha \beta}
= 
g_{\alpha \beta} 
+ \omega_{\alpha \beta},\;\;\;
\omega_{\alpha \beta} = -\omega_{\beta \alpha},
\end{equation} 
the spinor transformation matrix $\hat{S}(\Lambda)$  reads 
\begin{equation}
\hat{S}(\Lambda) = e^{
\frac{1}{4} \omega_{\alpha \beta} \sigma^{\alpha \beta}}
\approx  1 + \frac{1}{4} \omega_{\alpha \beta} \sigma^{\alpha \beta}.
\end{equation}
If we choose
\begin{equation}
\omega_{\mu \nu}
= 2 ( a_\mu k_\nu - 
a_\nu k_\mu ),
\end{equation}
we find the following equation 
for the spinor $u_a$
\begin{equation}
u_a = u(\Lambda q_t, \Lambda s_q).
\end{equation}
Then, writing 
\begin{equation}
 \sigma^{\nu \mu} k_\nu 
 a_\mu u(q_t,s_q) 
 = 
 u(\Lambda q_t, \Lambda s_q) 
 - u(q_t, s_q), 
\end{equation}
and expanding the right-hand side in powers of $k$ through 
linear terms, we find
\begin{equation}
\begin{split}
  k_{\nu} \sigma^{\nu \mu} a_\mu u
  = & \frac{\partial u}{\partial p_t^\sigma}
        \Lambda^{\sigma\rho} p_{t,\rho} + \frac{\partial u}{\partial s^\sigma}
        \Lambda^{\sigma\rho} s_\rho,
        \\
\frac{\partial u}{\partial p_t^\sigma}
        \Lambda^{\sigma\rho} p_{t,\rho}
  = & 2 a_\mu \left( p_t^{\mu} k^{\sigma}
        \frac{\partial}{\partial p_t^{\sigma}} - (p_t k) \frac{\partial}{\partial
        p_{t,\mu}} \right) u = d_t \, a_\mu L_t^\mu\, u ,
        \\
\frac{\partial u}{\partial s^\sigma} \Lambda^{\sigma\rho} s_\rho
   = & 2  a_\mu \left( s^{\mu} k^{\sigma}
  \frac{\partial}{\partial s^{\sigma}} - (s k) \frac{\partial}{\partial
  s_{\mu}} \right) u = d_t\, a_\mu S_t^\mu \, u, 
  \label{eq:LSdef}
\end{split}
\end{equation}
where $d_t=2(p_t k)$ is the denominator of the top propagator without the subleading
$k^2=\lambda^2$ term.
Eqs~(\ref{eq:LSdef}) implicitly define the operators
$L_t$ and $S_t$ as given by
\begin{equation}
    L_t^\mu=\frac{2}{d_t}\left(p_t^\mu\,k^\nu \frac{\partial}{\partial p_t^\nu}-(p_t k) \, \frac{\partial}{\partial p_{t,\mu}}\right), \, \quad\quad
    S_t^\mu=\frac{2}{d_t}\left(s^\mu\,k^\nu \frac{\partial}{\partial s^\nu}-(s k) \,\frac{\partial}{\partial s_{\mu}}\right).
\label{eq:LSdef1}
\end{equation}

Since the vector $a$ is arbitrary,  from  
eq.~(\ref{eq:LSdef}) we infer the 
following result
\begin{equation}
    k_\nu\sigma^{\nu\mu} u= d_t(L^\mu_t+S_t^\mu ) u.
    \label{eq:sigmaLS}
\end{equation}
Next, using the definition  of  the current $J_t$ in eq.~(\ref{eq:JtJb}),
and discarding terms of order $k$, we write the amplitude as
\begin{equation}
  \mathcal{R}_{P,{\rm f}}^\mu
  =  J^\mu_t  \bar{u} (q_t, s_q)
   N (q_t + k, q_i) u (q_i)
  + [(L_t^\mu  + S_t^\mu ) \bar{u} (p_t, s)] N u (p_i).
\end{equation}
We note that if  the arguments of  the function $N$ are not written explicitly,  it  is to be  
understood as $N(p_t,p_i)$.
To simplify the leading term, we write 
\begin{equation}
J_t^\mu  \bar{u} (q_t, s_q)
   N (q_t + k, q_i) u (q_i)=J_t^\mu  \bar{u} (q_t, s_q)
   N (q_t , q_i) u (q_i)+
   \frac{2p_t^\mu}{d_t} \bar{u}(p_t,s)\;  k^\alpha  \frac{\partial N}{\partial p_t^\alpha} u(p_i).
 \label{eq:Nexpansion}
 \end{equation}

As stated earlier, the function $N(q_t+k,q_i)$ is a well-defined function of its arguments and can be constructed  from  Feynman graphs. It is not uniquely  defined, however, if  momentum conservation is violated, and this is exactly what happens in eq.~(\ref{eq:Nexpansion}) both  in the leading 
term $N(q_t,q_i)$ and when derivative with respect to $p_t$  in the last term is taken.  To interpret this equation, we need to  assume that $N$ is extended in some  way 
to account for the  momentum non-conservation.  The ambiguity introduced by such arbitrariness must cancel in 
the end, since it was not present in the initial formula.
We will see later that this, in fact, 
is the case. 
We finally write 
 \begin{equation}
  \mathcal{R}_{P,{\rm f}}^\mu
  =  J^\mu_t \mathcal{B}_P(s_q,q) + 
  \bar{u}(p_t,s) \left[\left(L^\mu_t + \frac{\partial}{\partial p_{t, \mu}}
    \right)N\right]u(p_i)
  + [(L_t^\mu  + S_t^\mu ) \bar{u} (p_t, s)] N u (p_i),
\end{equation}
where
\begin{equation}
   \mathcal{B}_P(s_q,q)=\bar{u}(q_t,s_q) N(q_t,q_i) u(q_i).
\end{equation}

A similar calculation  can be performed  for the radiation off  the $b$-quark in the initial state. We obtain
\begin{equation}
  \mathcal{R}_{P,{\rm i}}^\mu =
J_i^\mu  \mathcal{B}_P(s_q,q)
  + \bar{u}(p_t,s) \left[\left(-L_i^\mu  + \frac{\partial}{\partial p_{i, \mu} }\right)N \right]u(p_i) - \bar{u} (p_t, s) [ L^\mu_i u (p_i)],
\end{equation}
where $J_i$ and $L_i$ are defined in eqs~(\ref{eq:JtJb}) and~(\ref{eq:Lbdef}).
Unlike the case of radiation off the top quark, no term analogous to  the $S_t$ operator
arises here, since the $u(p_i)$ spinor is a helicity eigenstate, and helicity is Lorentz invariant.
We thus find
\begin{equation}
\begin{split}
  \mathcal{R}_{P,{\rm f}}+\mathcal{R}_{P,{\rm i}}
  = & 
      J\, \mathcal{B}_P(s_q,q)
      +(L_t+S_t-L_i)\, [{\bar u}(p_t,s)N u(p_i)] 
      \\
  +& {\bar u}(p_t,s)\left[\left(\frac{\partial}{\partial p_t}+\frac{\partial}{\partial p_i} \right)
      N \right] u(p_i).
      \label{eqb27}
      \end{split}
\end{equation}
where, as usual, $J =J_t+J_i$.

We note that all   terms that appear in  the first line 
in eq.~(\ref{eqb27})
vanish if we multiply the equation by
$k^\mu$, while the term on the
second line does not.\footnote{In fact it does vanish in the  single top production case. It does not necessarily vanish if we consider some associated production process, and we prefer to keep  the discussion general.}
On the other hand, since this term is non-singular in the soft limit, its lack of transversality  must be compensated by non-singular contributions caused by the  radiation from  internal lines, that must have the form
\begin{equation}
    {\cal R}_{P,\rm int}^\mu = 
  \bar{u}(p_t,s)  N^\mu_{\tmop{reg}} u(p_i).
\end{equation}
Current conservation  implies 
\begin{equation}
   \left[\mathcal{R}_{P,\rm f}+\mathcal{R}_{P,{\rm i}} + 
  {\mathcal R}_{P,\rm int}
  \right] \cdot k
= {\bar u}(p_t,s)\, k \cdot  \left[\frac{\partial N}{\partial p_t}
  +\frac{\partial N}{\partial p_i} +  N_{\tmop{reg}}\right]u(p_i)
  = 0.
\end{equation}
In order for this equation to hold for any value of $k$ we 
therefore must have
\begin{equation}
  N_{\tmop{reg}} = -\left( \frac{\partial N}{\partial p_t} +
    \frac{\partial N}{\partial p_{i}}\right).
  \label{eq:Nreg}
\end{equation}
Thus, the full 
result for gluon emission in production reads
\begin{equation}
  \mathcal{R}_P=  \mathcal{R}_{P,{\rm f}}+\mathcal{R}_{P,{\rm i}}
  + \mathcal{R}_{P,{\rm int}}=
   J\, \mathcal{B}_P(s_q,q)
      +(L_t+S_t-L_i)\, \mathcal{B}_P(s).
\end{equation}

We now introduce the  
mapping from  $q$- to  $p$-momenta. We employ the mapping already used in ref.~\cite{Makarov:2023ttq}, and discussed at length near eq.~(\ref{eq:momentamapping}).
Since the mass of the top is not changed by the mapping, it must be possible to write it as a Lorentz transformation $\Lambda$  which is given in eq.~(\ref{eq:Lambdak}).
In the present context, we should remember that we also need a transformation for $s_q$, that can be conveniently chosen to be given by the same Lorentz transformation, so that the identities $s_q^2=s^2$ and $q_t\cdot s_q=p_t\cdot s$ hold. For convenience we
report here the  complete mapping transformation: 
\begin{equation}
\begin{split}
  q_t & =  \Lambda p_t 
 =   p_t-k + \frac{(p_t k)}{(p_t p_d)} p_d, \quad
  q_d= p_d -\frac{(p_t k)}{(p_t p_d)} p_d, \\
    s_q & = \Lambda s = 
s +\frac{(s k)}{(p_t p_d)}p_d - \frac{(s p_d)}{(p_t p_d)}k.
  \label{eq:AmappingP}
  \end{split}
\end{equation}
We  recall that also the decay momenta must change, since the top quark momentum  has changed. However, since this  change  is 
the  Lorentz transformation $\Lambda$, the decay amplitude does not change.
Our final result is then
\begin{equation}
  \mathcal{R}_P^\mu = J^\mu \mathcal{B}_P(s,p) + 
   D^\mu_{P,r} \mathcal{B}_P(s,p),
   \label{eq:RPamp1}
\end{equation}
where
\begin{equation}
\begin{split} 
& D_{P,r}=J D_{\rm rec}+(L_t+S_t-L_i),
\\
& D_{\rm rec} = \left(-k+\frac{(p_t k)}{(p_t p_d)}p_d\right) \cdot \frac{\partial}{\partial p_t }
                    -\frac{(p_t k)}{(p_t p_d)}p_d \cdot \frac{\partial}{\partial p_d}
                    +  \left( \frac{(k s)}{(p_t p_d)}p_d - \frac{(p_d s)}{(p_t p_d)}k \right) \cdot \frac{\partial}{\partial s}.
                    \end{split}
\end{equation}
$D_{\rm rec}$ is  the differential operator associated with the momenta and spin mappings, and it can be immediately read out of eq.~(\ref{eq:AmappingP}).
It is straightforward to verify  that $D_{P,r}$
preserves physical conditions, such 
as the momentum conservation, the on-shell conditions and the spin transversality condition, 
\begin{align}
D^\mu_{P,r}\,(p_i+p_u-p_t-p_d)^\nu &=0,
&D^\mu_{P,r}\, p_i^2 &=0,
&D^\mu_{P,r}\, p_d^2  &=0,
\nonumber\\
D^\mu_{P,r}\, p_t^2 &=0, &D^\mu_{P,r}\, s\cdot p_t  &=0,
&D^\mu_{P,r}\, s^2 &=0.
\label{eq.a43a}
\end{align}
Thus, eq.~(\ref{eq:RPamp1}) depends upon $\mathcal{B}_P(s,p)$ evaluated with
momenta and spin satisfying  the physical conditions, since the
derivative acts in a direction tangent to the manifold where the Born
amplitude is unambiguously defined.
\\

To obtain the full amplitude for the top production and decay process, 
we should  multiply eq.~(\ref{eq:RPamp1}) 
with the decay amplitude. Since, as discussed earlier, we can assume that the momentum mapping satisfies the equation
\begin{equation}
  {\cal B}_D(s_q,q)={\cal B}_D(s,p),
\end{equation}
the full amplitude for the production and decay reads
\begin{equation}
\mathcal{R}_{PD}^\mu=  \sum_{\lambda_s=\pm 1}\left[ J^\mu \mathcal{B}_P(s,\lambda_s)\mathcal{B}_D(s,\lambda_s) + \mathcal{B}_D(s,\lambda_s) D^\mu_{P,r}
  \mathcal{B}_P(s,\lambda_s)\right].
\end{equation}
As we explained earlier,  if we 
choose $s=s_D$ only  contribution with $\lambda_s=+1$ survives in the sum.  Thus, upon squaring the above formula we arrive at 
\begin{equation}
-g_{\mu \nu} \mathcal{R}_{PD}^\mu
\mathcal{R}_{PD}^{\nu,+}
=
|\mathcal{R}_{PD}^\mu|^2 =|\mathcal{B}_D|^2\left\{ - J^2 |\mathcal{B}_P(s)|^2 -J \cdot D_{P,r}|\mathcal{B}_P(s)|^2 \right\}_{s=s_D}.
\label{eq:sqampreal}
\end{equation}
The $k$-dependence is exposed in the above formula and, after momenta redefinitions, integration over gluon momentum factorises from the rest of phase space. The needed integrals in $k$ can be found in ref.~\cite{Makarov:2023ttq}.  
The
result for the linear term in $\lambda$ arising from the integration is given by
\newcommand{\normint}{\beta}
\begin{equation}
{\cal T}_\lambda\left[|{R_{PD}^\mu}|^2\right]
= -|\mathcal{B}_{PD}|^2{\cal T}_\lambda \left[\int \frac{\mathd^3 k}{2k^0(2\pi)^3} \; J^2   \; \right]+|\mathcal{B}_D|^2 \left[\tilde{D}_{P,r} |\mathcal{B}_P(s)|^2\right]_{s=s_D} ,
 \label{eq:Arp}
\end{equation}
where 
\begin{equation}
\tilde{D}_{P,r}= - {\cal T}_\lambda
\left[\int \frac{\mathd^3 k}{2k^0(2\pi)^3} \; J \cdot D_{P,r}  \right]. 
\end{equation}
We will not show the result of the integration of the first term in the above equation because, as we will see later, it can be combined with other contributions and argued to cancel in a way similar to what was found in ref.~\cite{Makarov:2023ttq}.
Computing the required integral explicitly for the second term in 
eq.~(\ref{eq:Arp}),  we find
\begin{equation}
\begin{split}
\tilde{D}_{P,r} = & - \normint\Bigg[
      \frac{(p_i s)(2 \mt^2 p_d-(p_t p_d) p_t) - (p_d s)(2 \mt^2 p_i-(p_t p_i)p_t)}
      {(p_t p_d)(p_t p_i)} \cdot \frac{\partial}{\partial s}
      \\
  +&  \left(\frac{\mt^2}{(p_t p_d)} p_d - \frac{\mt^2}{(p_t p_i)} p_i
      \right) \cdot \frac{\partial}{\partial p_t} -
      \frac{\mt^2}{(p_t p_i)} p_i \cdot \frac{\partial}{\partial p_i}
      - \frac{\mt^2}{(p_t p_d)} p_d \cdot \frac{\partial}{\partial p_d}
      \Bigg],
      \label{eq:ArpdD}
\end{split}
\end{equation}
where
\begin{equation}
  \normint = \frac{1}{2(2\pi)^2} \frac{\lambda\pi}{\mt}.
\end{equation}

\subsection*{Virtual corrections in  production }
We continue with the discussion 
of the virtual correction to the 
heavy line in the production subprocess. 
We write it as follows 
\begin{equation}
  \mathcal{V}_P  =  \int \frac{\mathd^4 k}{(2 \pi)^4} 
  \frac{-i}{k^2 - \lambda^2  + i \epsilon} \; F_{VP}(k,...),
  \end{equation}
where 
\begin{equation}
\begin{split}
F_{VP} & = \left[ \bar{u} (p_t, s) \gamma^{\mu} \frac{{\slashed p}_t
  + {\slashed k} + \mt}{(p_t + k)^2 - \mt^2 + i \epsilon} N (p_t + k, p_i + k)
                      \frac{{\slashed p}_i + {\slashed k}}{(p_i + k)^2 + i \epsilon}
                      \gamma_{\mu} u (p_i)
                      \right.
                      \\
   & + \left.  \bar{u} (p_t, s) \gamma^{\mu} \frac{{\slashed p}_t + {\slashed k} + \mt}{(p_t
  + k)^2 - \mt^2 + i \epsilon} N_{\mu} u (p_i) + \bar{u} (p_t, s) N_{\mu} 
  \frac{{\slashed p}_i + {\slashed k}}{(p_i + k)^2 + i \epsilon} \gamma^{\mu} u (p_i)
  \right].
\end{split}
\label{eq.a52a}
\end{equation}
The  first line provides 
a contribution where a virtual gluon is emitted by an incoming bottom and absorbed by the outgoing top quark,  
and the terms in the second line 
describe contributions where virtual 
gluons are emitted by either bottom or top quarks  and  are absorbed by 
the internal lines of the diagrams. 
Potential contributions where gluons 
are emitted and absorbed by internal 
lines are not shown as they cannot 
produce ${\cal O}(\lambda)$ corrections 
\cite{Makarov:2023ttq}. 
Using the Dirac equations, and neglecting contributions that 
cannot produce ${\cal O}(\lambda)$ corrections, we rewrite the above expression as follows 
  \begin{equation}
  \begin{split}
F_{VP} & =  
  \bar{u} (p_t, s)  \frac{2 p_t^{\mu} + k^{\mu} + \sigma^{\mu \nu} k_{\nu}}{(p_t +
  k)^2 - \mt^2 + i \epsilon} N (p_t + k, p_i + k) \frac{2 p_{t,\mu} + k_{\mu} -
  \sigma_{\mu \rho} k^{\rho}}{(p_i + k)^2 + i \epsilon} u (p_i) \\
  &  +   \bar{u} (p_t, s) \frac{2 p_t^{\mu}}{(p_t + k)^2 - \mt^2 + i
  \epsilon} N_{\mu} u (p_i) + \bar{u} (p_t, s) N_{\mu}  \frac{2
  p_i^{\mu}}{(p_i + k)^2 + i \epsilon} u (p_i) \,.
  \end{split}
\end{equation}
Using again eq.~(\ref{eq:Nreg}) and eq.~(\ref{eq:sigmaLS})
we arrive at
\begin{equation}
\begin{split}
  {\cal V}_P
  =& \int \frac{\mathd^4 k}{(2\pi)^4}\frac{-i}{k^2 - \lambda^2 + i\epsilon}\Bigg\{J_t \cdot
      J_{i} \;  {\cal B}_P(s) + [(J_i \cdot (L_t+S_t)+J_t \cdot L_{i} )
      \mathcal{B}_P(s)]
   \\  
             -& \bar{u}(p_t,s)\left(J_t \cdot \frac{\partial N}{\partial p_t}
                 +J_i \cdot \frac{\partial N }{\partial p_i}\right)u(p_i)\Bigg\},
                 \label{eq:VP}
\end{split}
\end{equation}
where the currents $J_t$ and $J_i$ have now changed appropriately for eq.~(\ref{eq.a52a}),
and the definitions of  $L_t$, $S_t$ and $L_i$ are given in 
eqs~(\ref{eq:LSdef1}) and (\ref{eq:Lbdef}) except that  the newly defined denominators $d_t,d_i$ should be used there. 

 The second term on the right hand side of  eq.~(\ref{eq:VP}) is not a total derivative. To remedy this,  we assume that spinors can be written as  functions of their momentum alone.\footnote{This
choice does not affect the $L_t+S_t$ and the $L_i$ derivatives, since
they act as  Lorentz transformations on the argument of the spinor,
leaving the mass and the constraints on the spin parameter $s$ unchanged.}
To do this, in the 
expression for  $u(p_t,s,\lambda_s)$ we systematically replace the mass
$\mt$ 
with $\sqrt{p_t^2}$. 
This implies the following modification in the density matrix  
\begin{equation}
\sum_{\lambda_s = \pm 1}  u(p_t,s,\lambda_s)\bar{u}(p_t,s,\lambda_s)=(\slashed{p}_t+\sqrt{p_t^2})\,. \label{eq:uubarchoice}
\end{equation}

The replacement 
$m_t \to \sqrt{p_t^2}$
is important for simplifying 
eq.~(\ref{eq:VP}),
because it  implies 
\begin{equation}
p_t \cdot \frac{\partial}{\partial p_t} u(p_t.s)=\frac{1}{2}  u(p_t.s).
\end{equation}
This result easily follows from the fact that the mass dimension of a spinor is $1/2$ 
and that once the mass is eliminated in favour of $\sqrt{p_t^2}$, 
$p_t$ becomes the only mass scale that appears in the formula for the spinor. Hence, we find 
\begin{equation}
\begin{split}
&  \bar{u}(p_t,s)\left(J_t \cdot \frac{\partial N}{\partial p_t}
    +J_i \cdot \frac{\partial N }{\partial p_i}\right)u(p_i)
\\
&   = \left(J_i \cdot \frac{\partial}{\partial p_i}+J_t \cdot \frac{\partial}{\partial p_t}\right) [ \bar{u}(p_t,s) N u(p_i) ]
- \left(\frac{1}{d_i}+\frac{1}{d_t}\right)\bar{u}(p_t,s) N u(p_i).
\end{split}
\end{equation}
Defining
\begin{equation}
  D_{P,v}= J_i \cdot (L_{t}+S_{t})+J_t \cdot L_{i} - J_i \cdot \frac{\partial}{\partial p_i}
  - J_t \cdot \frac{\partial}{\partial p_t},
  \label{eq:DPv}
\end{equation}
we write eq.~(\ref{eq:VP}) as
\begin{equation}
  {\cal V}_P
  = \int \frac{\mathd^4 k}{(2\pi)^4}\frac{-i}{k^2-\lambda^2 + i\epsilon}\left[\left(J_t \cdot
      J_{i}+\frac{1}{d_i}+\frac{1}{d_t}\right) {\cal B}_P(s) + D_{P,v}\mathcal{B}_P(s) \right].
              \label{eq:VP1}
\end{equation}
We note that the derivative $D_{P,v}$
violates the physicality constraint related to momentum  conservation; 
we will see that 
it is restored 
once the mass renormalisation is accounted for.  

Since the dependence on the gluon momentum $k$ is exposed in eq.~(\ref{eq:VP1}),
we can integrate over it. 
Similar to the discussion of the real-emission contribution, we leave terms 
proportional to ${\cal B}_P(s)$ as they are, since we
will argue later  that their cancellation is already demonstrated in ref.~\cite{Makarov:2023ttq}.
We obtain 
\begin{equation}
\begin{split}
  &  {\cal T}_\lambda[{\cal V}_P]
     = {\cal T}_\lambda\left[\int \frac{\mathd^4 k}{(2\pi)^4}\frac{-i}{k^2 - \lambda^2 +i\epsilon}
     \left( J_t \cdot J_{i}+\frac{1}{d_i}+\frac{1}{d_t}\right)\right]
     \mathcal{B}_P(s) + \tilde{D}_{P,v} \mathcal{B}_P(s),
      \\
  & \tilde{D}_{P,v}=
       \frac{\normint}{(p_t p_i)} \left( - (p_i s)\, p_t \cdot \frac{\partial}{\partial s}
      + \mt^2\, p_i \cdot \frac{\partial}{\partial p_i}
      +  \mt^2\, p_i \cdot \frac{\partial}{\partial p_t} \right).
\end{split}
\label{eq:VP2}
\end{equation}
We then multiply eq.~(\ref{eq:VP2}) by the decay amplitude, set $s=s_D$ to
get rid of the spin summation, compute  the interference with the Born
amplitude, and finally obtain
\begin{equation}
\begin{split}
2 |\mathcal{B}_{PD}|^2
{\cal T}_\lambda
& 
\left[\int \frac{\mathd^4 k}{(2\pi)^4}\frac{-i}{k^2- \lambda^2 +i\epsilon}
     \left  ( J_t \cdot J_{i}+\frac{1}{d_i}+\frac{1}{d_t}\right)\right]
  \\  
  & +
  |\mathcal{B}_D|^2 \left[\tilde{D}_{P,v}|\mathcal{B}_P(s)|^2
  \right]_{s=s_D}. 
  \end{split}
  \label{eq:Avp}
\end{equation}

\subsection*{Real corrections  in decay}
We can describe radiation in the top quark  decay following the approach 
used in the discussion of  the production 
process. 
The only essential difference is in the  
momentum mapping, that
we choose to coincide with the one discussed in ref.~\cite{Makarov:2023ttq}. 
Details are given in  eq.~(\ref{eq:mapdec}).
Both the initial top quark momentum and its spin vector are not 
affected by the mapping. 
The decay amplitude expanded through linear terms in the gluon momentum reads 
\begin{equation}
  {\cal R}^\mu_D(s)=J^\mu {\cal B}_D(s)+D_{D,r}^\mu {\cal B}_D(s),
  \quad\quad
  D_{D,r}^\mu =   J^\mu D_{\rm rec} - (L_t^\mu+S_t^\mu-L_f^\mu),
\end{equation}
where the currents are given in eq.~(\ref{eq:Jdec}).
The differential operator associated with the mapping  can be immediately read out of eq.~(\ref{eq:mapdec}).
It reads 
  \begin{equation}
  D_{\rm rec}
  =\left(-k+\frac{(p_f k)}{(p_2 p_f)}p_2\right) \cdot
      \frac{\partial}{\partial p_f} - \frac{(p_f k)}{(p_2 p_f)}  p_2 \cdot \frac{\partial}{\partial p_2}.
\end{equation}
The operators $L_t$ and $S_t$ are defined in eq.~(\ref{eq:LSdef}) but  now $d_t=-2 (p_t k)$ has to be used there. The definition of $L_f$ is the same as the one for $L_i$ after the replacement $i\rightarrow f$ is performed.
The operator $D_{D,r}$ is easily seen to preserve the physicality conditions
for the decay.
Proceeding as for eq.~(\ref{eq:sqampreal}), we get
\begin{equation}
|\mathcal{R}_{PD}^\mu|^2 =|\mathcal{B}_P|^2\left\{ - J^2 |\mathcal{B}_D(s)|^2 -J \cdot D_{D,r}|\mathcal{B}_D(s)|^2 \right\}_{s=s_P}.
\label{eq:sqamprealdec}
\end{equation}
After the momentum mapping, the integration over  $k$  factorises and can be performed. Defining
\begin{equation}
\begin{split}
  \tilde{D}_{D,r}=& {\cal T}_\lambda\left[\int \frac{\mathd^3 k}{2k^0(2\pi)^4} (-J \cdot D_{D,r})\right]
  = \normint \Bigg[ \frac{(p_fs)}{(p_tp_f)} p_t \cdot  \frac{\partial}{\partial s}
      \\
  & +  \frac{(p_t p_f)}{(p_2p_f)} p_2 \cdot \left( \frac{\partial}{\partial p_2}- \frac{\partial}{\partial p_f} \right)+ \left(p_t -  \frac{\mt^2}{(p_tp_f)} p_f \right) \cdot \left( \frac{\partial}{\partial p_f}+ \frac{\partial}{\partial p_t}\right) \Bigg]\,,
\end{split}
\end{equation}
we find for the total amplitude squared
\begin{equation}\mathcal{T}_\lambda \left[
|\mathcal{R}_{PD}|^2 \right]=|\mathcal{B}_{PD}|^2{\cal T}_\lambda\left[\int \frac{\mathd^3 k}{2k^0(2\pi)^4}(-J^2)\right] + |\mathcal{B}_P|^2 \left[\tilde{D}_{D,r}|\mathcal{B}_D(s)|^2\right] _{s=s_P}.\label{eq:Ard}
\end{equation}

\subsection*{Virtual corrections in decay}
The calculation of the virtual corrections in the decay process closely follows  the production case, and the result can be obtained from
eq.~(\ref{eq:VP2}) after the substitution $i \to f$. We obtain
\begin{equation}
\begin{split}
    {\cal T}_\lambda[{\cal V}_D]
     &= {\cal T}_\lambda\left[\int \frac{\mathd^4 k}{(2\pi)^4}\frac{-i}{k^2 - \lambda^2 + i\epsilon}
     \left( J_t \cdot J_{f}+\frac{1}{d_f}+\frac{1}{d_t}\right)\right]
     \mathcal{B}_D(s) + \tilde{D}_{D,v} \mathcal{B}_D(s),
     \\
   \tilde{D}_{D,v} &=
      \frac{\normint}{(p_t p_f)} \left( - (p_f s)\, p_t \cdot \frac{\partial}{\partial s}
      + \mt^2 \, p_f \cdot \frac{\partial}{\partial p_f}
      + \mt^2 \, p_f \cdot \frac{\partial}{\partial p_t}\right).
\end{split}
\end{equation}
The final contribution is
\begin{equation}
\begin{split}
2 |\mathcal{B}_{PD}|^2
{\cal T}_\lambda &\left[ \int \frac{\mathd^4 k}{(2\pi)^4}\frac{-i}{k^2 - \lambda^2 + i\epsilon}
     \left( J_t \cdot J_{f}+\frac{1}{d_f}+\frac{1}{d_t}\right)\right]
   \\
     & +
  |\mathcal{B}_P|^2 \left[\tilde{D}_{D,v}|\mathcal{B}_D(s)|^2\right]_{s=s_P}.
  \end{split}
  \label{eq:Avd}
\end{equation}

\subsection*{Top quark  self-energy contribution}

\begin{figure}
  \begin{center}
    \includegraphics[width=0.6\textwidth]{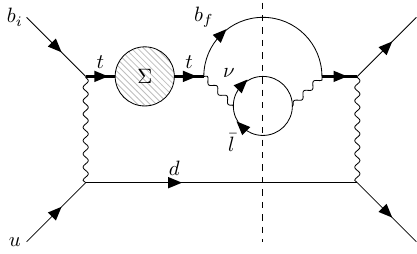}
    \caption{The graph with  the top quark self-energy insertion. The complex conjugate diagram 
      should also be added.}
    \label{fig:Selfen} 
  \end{center}
\end{figure}

Finally, we need to account for the  top quark self-energy correction shown in Fig.~{\ref{fig:Selfen}}.
Since we  perform the  calculation 
in a short-distance scheme
(e.g. the $\overline{\rm MS}$ scheme), no ${\cal O}(\lambda)$ correction can arise from the  mass counter-terms, and we do not need to account for them.
Following the discussion  in  section~7 in ref.~\cite{Makarov:2023ttq}, we find that 
the term of order $\lambda$ from the self-energy
insertion  to the left of the cut line in Fig.~\ref{fig:Selfen} reads
\begin{equation} \Sigma =\kappa  \left[ -
   \frac{1}{4 \mt} (p_t^2 - \mt^2) - ({\slashed p_t} - \mt) + \mt \right].
\end{equation}
where 
\begin{equation}
\kappa=4\pi \as \Cf \beta=\frac{\as\Cf}{2\pi}\,\frac{\pi \lambda}{\mt}.
\end{equation}
The product  of the denominators of the top propagators in Fig.~\ref{fig:Selfen}
yields
\begin{equation}  \frac{1}{(p_t^2 - \mt^2)^2 + (\mt \Gammat )^2}  \frac{1}{(p_t^2 - \mt^2)
   + i \mt \Gammat }.
\end{equation}
When this expression is combined with the complex conjugate of Fig.~\ref{fig:Selfen},
it becomes
\begin{equation} \frac{2 (p_t^2 - \mt^2)}{((p_t^2 - \mt^2)^2 + (\mt \Gammat )^2)^2} = -
   \frac{\partial}{\partial p_t^2} \frac{1}{(p_t^2 - \mt^2)^2 + ( \mt \Gammat)^2}
   \approx - \frac{\pi}{ \mt \Gammat} \delta' (p_t^2 - \mt^2), 
\end{equation}
where 
$\delta'(p_t^2-\mt^2)$  is the derivative of a delta function with respect to $p_t^2$.
Thus,  the net effect of the self-energy correction
amounts to the following replacement in the Born cross section
\begin{equation}
\begin{split}
& i (\slashed{p}_t+\mt) \delta(p_t^2-\mt^2)
  \rightarrow 
   -i (\slashed{p}_t+\mt)i \Sigma i (\slashed{p}_t+\mt) \delta'(p_t^2-\mt^2) 
   \\
& \phantom{aaaaaaaaaa}  = i \kappa \left[\frac{3}{2}(\slashed{p}_t+\mt)-\mt\right]\delta(p_t^2-\mt^2)
   +i \kappa(\slashed{p}_t+\mt)2m_t^2 \delta'(p_t^2-\mt^2), 
\end{split}
\label{eq:selfen}
\end{equation}
where  the relation $\delta' (x) x = - \delta (x)$ was used.

In the self-energy computation, one needs to consider slightly off-shell momenta of the top
quark. This means that we cannot set $p_t^2=\mt^2$ in the factor involving the derivative of the
delta function. We thus rewrite this term as follows
\begin{equation}
  \slashed{p}_t+\mt = (\slashed{p}_t+\sqrt{p_t^2})+ \frac{\mt^2-p_t^2}{\mt+\sqrt{p_t^2}}.
\end{equation}
Inserting this identity in eq.~(\ref{eq:selfen}) we obtain
\begin{equation}
   (\slashed{p}_t+\mt) \delta(p_t^2-\mt^2) \rightarrow   \kappa \frac{3}{2}(\slashed{p}_t+\mt)\delta(p_t^2-\mt^2)
   + \kappa(\slashed{p}_t+\sqrt{p_t^2})2m_t^2 \delta'(p_t^2-\mt^2). \label{eq:selfen1}
\end{equation}
The first term in the above equation leads 
to a term proportional to $|{\cal B}_{PD}|^2$, and it can be set aside with all
other terms of this form.
Using for consistency eq.~(\ref{eq:uubarchoice}), we obtain for the second term
\begin{equation}
  |\mathcal{B}_{PD}|^2 (2 \mt^2 \kappa \delta' (p_t^2 - \mt^2) )=
   |\mathcal{B}_{PD}|^2 \left[\delta \left( p_t^2 - \mt^2 \left( 1 - 2\kappa \right) \right) -
   \delta (p_t^2 - \mt^2)\right] . 
\end{equation}
In order to compute the result we need to perform a change of variables in  the first term. We first relabel all the $p$'s into $q$'s and $s$ into $s_q$.
In particular
\begin{equation}
  \delta \left( p_t^2 - m_t^2 \left( 1 - 2\kappa \right) \right)\rightarrow
  \delta \left( q_t^2 - m_t^2 \left( 1 - 2\kappa \right) \right).
\end{equation}
Then we recall that in the unpolarised case \cite{Makarov:2023ttq} two different mappings were used for single-top production 
process and 
for top 
decay. The momenta and spin transformations for the decay are given in eq.~(\ref{eq5.1}), and we report them here for convenience
\begin{equation}
  q_t = p_t (1 - \kappa),
  \;\;\;\; q_f =  p_f - \kappa p_t + \kappa \frac{(p_f p)}{(p_f p_2)} p_2,\;\;\;
  q_2  =  p_2 \left( 1 - \kappa \frac{(p_f p_t)}{(p_f p_2)} \right),\quad s_q=s.
\label{eq:mdec}
\end{equation}
Notice that the spin is unchanged, since the top momentum is only rescaled. For the production we use 
\begin{equation}
  q_t =  p_t - \kappa \frac{p_t^2}{(p_d p_t)}p_d,\;\;\;\;
  q_d  =  p_d \left( 1 + \kappa \frac{p_t^2}{(p_d p_t)} \right),\quad s_q=\Lambda_m s,
\label{eq:mprod}
\end{equation}
where $\Lambda_m=1+\kappa\omega_{td}$ (see eq.~(\ref{eq5.10})) and $\omega_{td}$ is  given in eq.~(\ref{eq:omegaxy}). The transformation for $s_q$ can be seen to satisfy the condition
$s_q\cdot q_t=s\cdot p_t$ and $s_q^2=s^2$,
as will become clear
in the following.
We should modify these
transformations in such a way that the
top quark momentum  transforms in the same way in both production \emph{and} decay processes. This is achieved
by using the fact that $\Lambda_m$ is a Lorentz transformation, and that the top momentum mapping can be written as the product of a rescaling times $\Lambda_m$. It is sufficient then to apply $\Lambda_m$ also to all decay products. In summary, we write
\begin{equation}
\begin{split}
  q_t & = (1-\kappa) \Lambda_m p_t, \;\;\;\;\;\;
   q_d  =  p_d \left( 1 + \kappa \frac{p_t^2}{(p_d p_t)} \right), \;\;\;\;\;
  s_q   =  \Lambda_m s,
 \\
  q_f & =  \Lambda_m \left ( p_f - \kappa p_t + \kappa \frac{(p_f p_t)}{(p_f p_2)} p_2 \right ), \;\;\;\;\;\
  q_2   = \left( 1 - \kappa \frac{(p_f p_t)}{(p_f  p_2)} \right) \Lambda_m p_2,
  \;\;\;\;
    q_1  =  \Lambda_m p_1,
  \end{split}
\label{eq:massmap}
\end{equation}
where only ${\cal O}(\kappa)$ terms need to be retained on the right hand sides of the above equations. Notice that the first line is just a rewriting of eq.~(\ref{eq:mprod}), while the second line is the  Lorentz transformation  applied to the  eq.~(\ref{eq:mdec}).
Momenta modifications 
induce changes in the 
squared amplitudes. Depending on whether momenta in the production or decay undergo these  transformations, 
we group such terms into modification of production and decay amplitudes and write 
\begin{equation}
   D^m \Big| \sum_{\pm 1} \mathcal{B}_D(\lambda_s)\mathcal{B}_P(\lambda_s) \Big|^2
  =   |\mathcal{B}_P|^2 [D^m|\mathcal{B}_D(s)|^2]_{s=s_P}
  +   |\mathcal{B}_D|^2 [D^m|\mathcal{B}_P(s)|^2]_{s=s_D}.
\end{equation}
The differential operator  $D^m$ is  associated with the transformations shown 
in eq.~(\ref{eq:massmap}).  It can be written as 
\begin{eqnarray}
  D^m|\mathcal{B}_D(s)|^2
  &=& \kappa\left[
      - p_t \cdot \frac{\partial}{\partial p_t}
      +\left(- p_t + \frac{(p_fp_t)}{(p_fp_2)} p_2\right) \cdot \frac{\partial}{\partial p_f}
      - \frac{(p_f p_t)}{(p_f p_2)} p_2 \cdot \frac{\partial}{\partial p_2}\right]|\mathcal{B}_D(s)|^2,
      \phantom{aaaaaa}
      \label{eq:selfencorrD}
  \\
  D^m|\mathcal{B}_P(s)|^2
  &=& \kappa\left[(-p_t+\omega_{td}p_t)\cdot\frac{\partial}{\partial p_t}
      +\frac{p_t^2}{(p_d p_t)} p_d \cdot \frac{\partial}{\partial p_d}
      + (\omega_{td} s)\cdot \frac{\partial}{\partial s}\right]|\mathcal{B}_P(s)|^2,
      \label{eq:selfencorrP}
\end{eqnarray}
where we have simply dropped from $D^m$ the derivatives with respect to variables not
contained in the corresponding amplitude and, in the case of the derivative of the decay
amplitude,  we have removed the Lorentz transformation, since it affects all the decay  momenta
and the spin  vector $s$, and the decay amplitude is Lorentz invariant.

\subsection*{Assembling everything}

We begin by considering the 
non-derivative terms  that appear in 
eqs (\ref{eq:Arp}, \ref{eq:Avp},
\ref{eq:Ard}, \ref{eq:Avd}, \ref{eq:selfencorrD}, \ref{eq:selfencorrP}). They arise from the dominant terms
in the cross sections, from Jacobians due to momenta  transformations   and in  the calculation of the virtual contributions in production and decay. Their calculation is straightforward, and can be carried out along the lines of ref.~\cite{Makarov:2023ttq}, where it has been  shown that they cancel.

We continue with terms  that contain derivatives with respect to momenta and 
spins of external particles.  Combining such 
contributions in the virtual corrections for production, eq.~(\ref{eq:Avp}), the real emission contributions in production, eq.~(\ref{eq:Arp}), and the corresponding part of the self-energy correction, eq.~(\ref{eq:selfencorrP}), we find
\begin{equation}
  |\mathcal{B}_D|^2\left[\left(D^m+\frac{\kappa}{\normint}[\tilde{D}_{P,v}+\tilde{D}_{P,r}]\right)|\mathcal{B}_P(s)|^2\right]_{s=s_D}.
  \label{eq:a84}
\end{equation}
It is a matter of simple algebra to verify that the derivative operator $D^m+\kappa/\beta \tilde{D}_{P,v}$,
with $D^m$ restricted to the production variables, and, independently, $\tilde{D}_{P,r}$ 
preserve  all physicality conditions, c.f. 
eq.~(\ref{eq.a43a}).
The result shown in eq.~(\ref{eq:a84}) becomes 
\begin{equation}
  |\mathcal{B}_D|^2 \kappa \Bigg[  \left\{s \cdot \Big ( 2 \omega_{ti} + 2 \omega_{dt}
    + \frac{2 m_t^2 (p_i p_d)}{(p_t \pbi)(p_t p_d) }  \omega_{i d}  \Big) \cdot
    \frac{\partial }{\partial s } \right\}|\mathcal{B}_P(s)|^2 \Bigg]_{s=s_D}.
\label{eqa.85}
\end{equation}
For the decay contribution, assembling eqs~(\ref{eq:Ard}), (\ref{eq:Avd}) and~(\ref{eq:selfencorrD}) we get
\begin{equation}
  |\mathcal{B}_P|^2\left[\left(D^m+\frac{\kappa}{\normint}[\tilde{D}_{D,v}+\tilde{D}_{D,r}]\right)|\mathcal{B}_D(s)|^2\right]_{s=s_P}=0.
  \label{eqa.86}
\end{equation}
The sum of eqs (\ref{eqa.85}) and (\ref{eqa.86})  agrees with  eq.~(\ref{eq6.1}).

\bibliographystyle{JHEP}
\bibliography{sintop}

\end{document}